\documentclass[useams,useamsmath,usenatbib]{mnras} \pdfminorversion=5
\usepackage{amsmath} \usepackage{setspace} \usepackage{pdfpages}
\graphicspath{{./images/}}

\def\ie{\,{\rm i.e.}\,} \def\eg{\,{\rm e.g.}\,}

\def\lambdar{{\sc lambdar}} \def\magphys{{\sc magphys}}   
\def\bbf{}

\begin{document}

\twocolumn

\title[GAMA: GSMF to $z=0.1$]{Galaxy And Mass Assembly (GAMA): The galaxy stellar mass function to $z=0.1$ from the r-band selected equatorial regions.}

%Authors {{{
\author[Wright et al.] {
A.H.~Wright$^{1,2}$\thanks{e-mail:awright@uni-bonn.de},
A.S.G.~Robotham$^{1}$,
S.P.~Driver$^{1,3}$,
M.~Alpaslan$^{4}$,
S.K.~Andrews$^{1}$,
\newauthor
I.K.~Baldry$^{5}$,
J.~Bland-Hawthorn$^{6}$
S.~Brough$^{7}$,
M.J.I.~Brown$^{8}$,
M.~Colless$^{9}$,
\newauthor
E.~da~Cunha$^{9}$,
L.J.M.~Davies$^{1}$,
Alister~W.~Graham$^{10}$,
B.W.~Holwerda$^{11}$,
\newauthor
A.M.~Hopkins$^{6}$,
P.R.~Kafle$^{1}$,
L.S.~Kelvin$^{5}$,
J.~Loveday$^{11}$,
S.J.~Maddox$^{13,14}$,
\newauthor
M.J.~Meyer$^{1}$,
A.J.~Moffett$^{6}$,
P.~Norberg$^{15}$,
S.~Phillipps$^{16}$,
K.~Rowlands$^{3}$,
\newauthor
E.N.~Taylor$^{10}$,
L.~Wang$^{17,18}$,
S.M.~Wilkins$^{12}$ \\
$^{1}$Argelander-Institut f{\"u}r Astronomie, Universit{\"a}t Bonn, Auf dem H{\"u}gel 71, 53121 Bonn, Germany \\
$^{2}$ICRAR, The University of Western Australia, 35 Stirling Highway, Crawley, WA 6009, Australia \\
$^{3}$SUPA, School of Physics \& Astronomy, University of St Andrews, North Haugh, St Andrews, KY16 9SS, UK \\
$^{4}$NASA Ames Research Center, N232, Moffett Field, Mountain View, CA 94035, US \\
$^{5}$Astrophysics Research Institute, Liverpool John Moores University, IC2, Liverpool Science Park, 146 Brownlow Hill, Liverpool, L3 5RF \\
$^{6}$Sydney Institute for Astronomy, School of Physics, University of Sydney, NSW 2006, Australia \\
$^{7}$Australian Astronomical Observatory, PO Box 915, North Ryde, NSW 1670, Australia \\
$^{8}$School of Physics and Astronomy, Monash University, Clayton, Victoria, 3800, Australia \\
$^{9}$Research School of Astronomy and Astrophysics, Australian National University, Canberra, ACT, 2611, Australia \\
$^{10}$Centre for Astrophysics and Supercomputing, Swinburne University of Technology, Victoria, 3122, Australia \\
$^{11}$Department of Physics and Astronomy, University of Louisville, Louisville KY 40292 USA \\
$^{12}$Astronomy Centre, University of Sussex, Falmer, Brighton, BN1 9QH, UK \\
$^{13}$SUPA, Institute for Astronomy, University of Edinburgh, Royal Observatory, Blackford Hill, Edinburgh EH9 3HJ, UK \\
$^{14}$School of Physics and Astronomy, Cardiff University, The Parade, Cardiff CF24 3AA, UK \\
$^{15}$ICC \& CEA, Department of Physics, Durham University, South Road, Durham DH1 3LE, UL \\
$^{16}$Astrophysics Group, School of Physics, University of Bristol, Tyndall Avenue, Bristol, BS8 1TL, UK \\
$^{17}$SRON Netherlands Institute for Space Research, Landleven 12, 9747 AD, Groningen, The Netherlands \\
$^{18}$Kapteyn Astronomical Institute, University of Groningen, Postbus 800, 9700 AV, Groningen, The Netherlands
}
\date{Accepted 2017 May 9. Received 2017 April 24; in original form 2017 January 18.}
\pubyear{2016} \volume{000} \pagerange{\pageref{firstpage}--\pageref{lastpage}}
%}}}
\maketitle \label{firstpage}
\begin{abstract}%{{{
{\bbf We derive the low redshift
galaxy stellar mass function (GSMF), inclusive of dust corrections, for the
equatorial Galaxy And Mass Assembly (GAMA) dataset covering 180 deg$^2$.
%This
%work is designed as an update of the analysis presented in \cite{Baldry2012},
%which used the preliminary GAMA I dataset from the first three years of the
%survey.
We construct the mass function using a density-corrected maximum
volume method, using masses corrected for the
impact of optically thick and thin dust. We explore the galactic bivariate brightness
plane ($M_\star-\mu$), demonstrating that surface
brightness effects do not systematically bias our mass function measurement above
10$^{7.5}$ M$_{\odot}$.  The galaxy distribution in the $M-\mu$-plane appears well bounded, indicating
that no substantial population of massive but diffuse or highly compact
galaxies are systematically missed due to the GAMA selection criteria.
The GSMF is {fit with} a double Schechter
function, with $\mathcal M^\star=10^{10.78\pm0.01\pm0.20}M_\odot$,
$\phi^\star_1=(2.93\pm0.40)\times10^{-3}h_{70}^3$Mpc$^{-3}$,
$\alpha_1=-0.62\pm0.03\pm0.15$,
$\phi^\star_2=(0.63\pm0.10)\times10^{-3}h_{70}^3$Mpc$^{-3}$, and
$\alpha_2=-1.50\pm0.01\pm0.15$. We find the equivalent faint end slope
as previously estimated using the GAMA-I sample, although we find a higher value of $\mathcal M^\star$.
Using the full GAMA-II sample, we are able to fit the mass function to masses as low as
$10^{7.5}$ $M_\odot$, and assess limits to $10^{6.5}$ $M_\odot$.
Combining GAMA-II with data from G10-COSMOS we are able to comment
qualitatively on the shape of the GSMF down to masses as low as $10^{6}$
$M_\odot$. Beyond the well known upturn seen in the GSMF at $10^{9.5}$ the
distribution appears to maintain a single power-law slope from $10^9$ to $10^{6.5}$.
We calculate the stellar mass density parameter given our best-estimate GSMF, finding
$\Omega_\star= 1.66^{+0.24}_{-0.23}\pm0.97 h^{-1}_{70} \times 10^{-3}$, inclusive of random and systematic
uncertainties.}
\end{abstract}
%}}}
%}}}

\section{Introduction}\label{sec: intro} %{{{
The Galaxy Stellar Mass Function
(GSMF; \citealt{Bell2003,Baldry2008,Baldry2012}) is arguably one of the most fundamental
measurements in extra-galactic astronomy. Its integral returns the density of
baryonic mass currently bound in stars (and hence the global efficiency of
star-formation) while the shape of the distribution describes the evolutionary
pathways which have shuffled matter from atomic to stellar form --- essentially
mergers building the high mass end of the GSMF ($M_\star \geq 10^{10.8}$)
while in-situ star-formation
fueled by gas accretion has built the low mass end \citep{Robotham2014}. Not
surprisingly the GSMF is also {\it the} key calibration for most galaxy
formation models that are carefully tuned to best reproduce the latest GSMF
measurement \citep{Schaye2015,Crain2015,Lacey2016,Gonzalez-Perez2014,Genel2014}.
In particular the comparison between observations of the GSMF and numerical
simulations of the dark-matter halo mass function have led directly to the
notion of feedback --- both AGN feedback at high mass (see, \eg
\citealt{Bower2006,Croton2006}) and supernova feedback at low mass
\citep{Efstathiou2000}. These are now core elements of semi-analytic
prescriptions used to populate the halos formed in purely dark-matter N-body
simulations \citep{Lacey2016,Gonzalez-Perez2014}.

Observationally the measurement of the GSMF has superseded the earlier focus on the
measurements of the galaxy luminosity function. Initially these were undertaken
in the optical and later at near-IR wavelengths, where the near-IR light was
shown to more closely trace the low mass stellar populations that dominate the
stellar mass repository. Near-IR is the best single-band proxy for stellar mass because
near-IR colours contain little information about mass-to-light variations. This conspires
to mean there is less scatter in near-IR single-band mass-to-light estimates compared to the
same proxies measured in the optical. Once multi-band optical and NIR data became ubiquitous, however, better
estimates could be obtained by making use of full SED colour information. Ultimately a
lot of information on optical mass-to-light is contained in the restframe g-r-i colours,
so surveys such as SDSS and GAMA could make estimates of stellar mass content that are
accurate within $<0.2$ dex \citep{Taylor2011}. Over the past two decades the ability to estimate
stellar mass has also become more established (see, \eg
\citealt{Bell2007,Kauffmann2003,Taylor2011}).  As a consequence effort has now
shifted from measuring galaxy luminosity functions to the GSMF. The most notable
measurements are those deriving from large redshift surveys, in particular the
2dF Galaxy Redshift Survey (2dFGRS; \citealt{Cole2001}), the Sloan Digital Sky
Survey (SDSS; \citealt{Bell2003,Baldry2008}), the Millennium Galaxy Catalogue
(MGC; \citealt{Driver2007}), and the Galaxy And Mass Assembly Survey (GAMA;
\citealt{Baldry2012}). In general there is a reasonable consensus with the
latest measurement from the GAMA team \citep{Baldry2012}, probing to a stellar
mass limit of $10^{8}$M$_{\odot}$.

However three key observational concerns remain: susceptibility to surface
brightness selection effects, the impact of dust attenuation, and the prospect
of a sharp upturn in the space density at very low stellar masses (i.e., below
the current observational mass limits). All three effects could potentially lead
to underestimating the GSMF and the corresponding stellar mass density. This is
particularly significant when looking to reconcile the current
stellar mass density with the integral of the cosmic star-formation history
(CSFH; see \citealt{Wilkins2008,Baldry2003}), where a significant discrepancy was seen.
In an
attempt to explain this discrepancy, some studies have invoked
either a top-heavy IMF (which produces more
luminosity per unit mass of stars; \citealt{Baldry2003}), a time varying IMF
\citep{Wilkins2008a,Ferreras2015,Gunawardhana2011}, distinct IMFs for bulge (closed-box star-formation with
a top-heavy IMF) and disc formation (infall star-formation with a standard
Chabrier-like IMF) as proposed by \cite{Lacey2016}, or an IMF with a larger fraction of returned mass
(\eg \citealt{Maraston2005}; see also \citealt{Madau2014}).
Additionally, the integrated cosmic star-formation history will tend to capture all star formation events
without consideration of dynamical interactions that deposit formed stars into the intra-halo medium (IHM).
This means that the integrated cosmic star-formation history naturally includes stellar material not currently
bound to observed galaxies. The combination of the CSFH and the GSMF measured across a broad redshift range
is therefore a powerful tool to constrain the IMF, feedback and extraneous material stripped from galaxies.

The first comprehensive measurements of the GSMF were made by \cite{Cole2001}.
This was based on the combination of spectroscopic measurements from the 2dFGRS
combined with photometric near-IR measurements from 2MASS. {\bbf Concurrently,
\cite{Kochanek2001} also used 2MASS to estimate the value of $\Omega_\star$ from
K-band luminosity function, although did not calculate the GSMF explicitly.}
\cite{Andreon2002} subsequently demonstrated that the shallow 2MASS survey misses dim galaxies entirely and
significantly underestimated the fluxes of late-type systems. Similarly the
later and larger studies based on SDSS and GAMA are both reliant on the
completeness of the spectroscopic input catalogues derived from (relatively)
shallow drift-scan SDSS imaging. \cite{Blanton2005} demonstrated, via adding
simulated galaxies to SDSS data, that incompleteness in the imaging and
spectroscopy can become severe for systems with average surface brightnesses of
$\mu_{50,r} \approx 23.5$mag/sq arcsec (see Figure $2$ of \citealt{Blanton2005},
and Figure $11$ of \citealt{Baldry2012}).  However one indication that the
surface brightness problem may not be overly severe comes from deep field
studies (see, \eg \citealt{Driver1999}), novel analysis methods designed to search for
low-surface brightness galaxies in wide-field imaging \citep{Williams2016},
and dedicated low-surface brightness
studies (see, \eg \citealt{Davies2016b,Geller2012}), which generally found that
large populations of low surface brightness systems do not contribute
significantly to the stellar mass density. Furthermore, attempts to correct
galaxy luminosity function estimates via a bivariate brightness analysis also
failed to find extensive populations of low surface brightness giant galaxies
(see, \eg \citealt{Cross2001,Driver2005}).

Dust attenuation has perhaps a more subtle effect. Generally dust will both
diminish and redden a galaxy's emission, and these two effects arguably cancel
---  the reduction in total light is compensated for by an increase in the
estimated mass-to-light ratio (see, \eg the vector shown in Figure 6 of
\citealt{Bell2003}, and Figure 11 of \citealt{Taylor2011}).  Strictly this is
only true in the optically thin case, as if no light from a particular region is
able to escape then the loss of flux cannot be recovered. The MGC team
\citep{Driver2007} attempted to quantify the impact of dust attenuation on
galaxy mass estimates by measuring the shift in the recovered ${\mathcal
M}^\star$-parameter of the optical $B$-band luminosity function with systemic
inclination. The implicit assumption was that, if dust attenuation is
significant, edge-on systems should be more attenuated than their face-on
counterparts. A significant $M_\star-cos(i)$ effect was seen \citep{Driver2007}
which, following extensive modelling using radiative transfer codes
\citep{Tuffs2004,Popescu2000}, suggested that the average face-on central
opacity of galaxy discs was $\tau_{v}=3.8$; \ie the centres of galaxies are
optically thick. The resulting impact, based on corrections using the radiative
transfer models, was to increase the estimate of the present day integrated
stellar mass density from $\sim5\%$ \citep{Baldry2008} to $\sim8\%$
\citep{Driver2007}. However significant concerns remain as to the validity of
adopting a constant central face-on opacity for all galaxy types. Indeed,
direct observations of galaxies have indicated that the intrinsic
nature of dust in galaxies is highly variable, depending on multiple factors
such as morphology and environment (see, \eg, \citealt{White2000,Keel2001,
Holwerda2005,Holwerda2013a,Holwerda2013b}).

Measurements of the GSMF to date reliably extend only to $10^8$M$_{\odot}$
whereas we have proof-of-existence of galaxies with masses as low as
$10^3$M$_{\odot}$ in the Local Group \citep{McConnachie2012}.  Hence there is
also some uncertainty as to whether an extrapolation of the GSMF from
$10^8$M$_{\odot}$ to $10^3$M$_{\odot}$ is valid. Recently the study by
\cite{Moffett2016}, where the stellar mass functions was divided by galaxy type,
showed two populations with very rapidly rising slopes at the mass-limit
boundary.

All three areas (surface brightness, dust attenuation, and low mass systems)
have the potential to bring into question the robustness of our current
estimates of the GSMF and the integrated cosmic stellar mass density. In this
paper we provide an updated GSMF, defined using the SDSS r-band,
for the completed Galaxy And Mass Assembly (GAMA;
\citealt{Driver2011,Liske2015}) survey equatorial fields.

In
Section 2, we introduce the GAMA--II sample which is approximately double the size
of the GAMA--I sample used in \cite{Baldry2012}, extending 0.4mag deeper (to
$r=19.8$mag) and over an expanded area of 180 sq deg. We also utilise the full
GAMA panchromatic imaging dataset \citep{Driver2016}, and photometry measured
consistently in all bandpasses from far-UV to far-IR \citep{Wright2016}. The
far-IR data from Herschel ATLAS \citep{Eales2010} in particular allow for full
SED modelling using codes such as \magphys\ \citep{daCunha2008,daCunha2011},
which accounts for dust attenuation and re-emission when calculating stellar masses.
In Section 3 we compare the stellar masses derived from optical
data using stellar template modelling \citep{Taylor2011} to those derived via
the full SED modelling from \magphys.  In Section 4, we derive our base
GSMF, incorporating density modelling of the GAMA volumes.  In Section 5 we
revert to a simpler empirical 1/V$_{\rm max}$ method applied in the bivariate brightness
plane to specifically explore the possible impact of surface brightness
selection bias. Finally in Section 6 we include similar photometric data from
the G10-COSMOS regions \citep{Davies2015a,Andrews2016}, fit with \magphys\
\citep{Driver2016} using high precision photometric redshifts from
\cite{Laigle2016}, to provide an
indication as to the possible form of the stellar mass function to very low
stellar masses ($10^6$M$_{\odot}$). We discuss our results in Section 7.
Throughout this work we use a standard concordance cosmology of $\Omega_M=0.3$,
$\Omega_{\Lambda}=0.7$, $H_o = 70$ kms$^{-1}$Mpc$^{-1}$, and $h_{70}=H_o/70$
kms$^{-1}$Mpc$^{-1}$. We implement a standard \cite{Chabrier2003} IMF, and all
magnitudes are presented in the AB system.
\vspace{15pt}
%}}}
\section{Data and Sample Definition}\label{sec: One} %{{{
%What is GAMA? Why is the survey useful for this task? {{{
The Galaxy And Mass Assembly (GAMA;
\citealt{Baldry2010,Driver2011,Liske2015,Hopkins2013}) survey is a large
multi-wavelength dataset built upon a spectroscopic campaign aimed at
measuring redshifts for galaxies with $r<19.8$ mag at $>98\%$ completeness \citep{Robotham2010}. The
survey's complementary multi-wavelength imaging is in 21 broadband photometric
filters \citep{Driver2016} spanning from the far-UV (FUV) to the far-IR (FIR).
%}}}
%Photometric Data {{{
Given this wealth of broadband imaging, we are able to calculate matched
photometry for the purposes of estimating galaxy stellar masses.  We use 21-band
photometry contained in the GAMA \lambdar\ Data Release (LDR), presented in
\cite{Wright2016}.  The LDR photometry is deblended matched aperture photometry accounting for each image's pixel resolution and point spread function.
Apertures used in \lambdar\ are defined using a mixture of source extractions on the SDSS r-band,
source extractions on the VISTA Z-band, and by-hand definitions using VISTA Z-band images.
Measurements are made for all images in the GAMA Panchromatic Data Release \citep{Driver2016}.

This photometric dataset is designed specifically for use in calculating
spectral energy distributions (SEDs), as the photometry and uncertainties are
consistently measured across all passbands. Furthermore, as the photometry
is matched aperture, there exists an estimate in every band for every object in
the sample, with a corresponding uncertainty (except, of course, where there is
no imaging data available due to coverage gaps). {\bbf For the calculation of relevant cosmological
distance parameters and redshift limits, fluxes have been appropriately k-corrected using
{\tt KCorrect} \citep{Blanton2007}, and redshifts have been flow-corrected using using the models
of \cite{Tonry2000} as described in \cite{Baldry2012}. }

%}}}
%}}}
%\section{SED fits with {\sc magphys} and {\sc interest}} %{{{
%Stellar Mass Estimates {{{
We calculate stellar masses for the LDR photometry using two independent
methods.  Firstly, we fit panchromatic SEDs to the full 21-band dataset using
the energy balance program \magphys\ \citep{daCunha2008,daCunha2011}. A full
description of the \magphys\ fits to the GAMA LDR is provided in
\cite{Driver2016b}. \magphys\ utilises information from the UV to the FIR to
estimate the total stellar mass of each galaxy from both visible and obscured
stars, assuming \cite{Bruzual2003} (BC03) models, a \cite{Chabrier2003} initial
mass function (IMF), and the \cite{Charlot2000} dust obscuration law. Secondly,
we use the measurement of \cite{Taylor2011} who estimated stellar masses by
fitting a comprehensive grid of SED templates to photometry from the SDSS
$u$-band to the VIKING $K_s$-band, applied to our updated LDR photometry.
Their technique uses stellar population
synthesis models with exponentially declining star-formation histories, without
bursts, and the same BC03 models and \cite{Chabrier2003} IMF as \magphys, but
uses a \cite{Calzetti2000} dust obscuration law.  In addition to this difference
in implemented dust obscuration law, the predominant differences between these
two methods are:
\begin{itemize}
\item the wider range of photometric filters (and energy balance) used in \magphys;
\item the incorporation of bursty star-formation histories in \magphys;
\item a sparser grid of star formation histories in \magphys.
\end{itemize}
For clarity, throughout this work we refer to stellar mass
estimates from \magphys, which utilise the full far-UV to far-IR bandpass, as `bolometric' masses,
and stellar masses from our stellar population synthesis templates, which are fit across the near-UV
to near-IR passbands, as `optical' masses.

%}}}
%Comparison between estimates {{{
Using these two methods, we check for systematic differences in our estimated
stellar masses. By comparing the two sets of mass estimates, we can
explore how our subsequent fits are systematically affected by our choice of
stellar mass estimation. In particular, an observed difference in the mass estimates (and GSMF fits)
can indicate the impact of optically thick dust on our masses (as \magphys\ includes consideration of optically
thick dust, whereas our optically estimated masses do not).

\begin{figure*} \centering
%\vspace{3in}
%\includegraphics[scale=0.7]{Ms_Comparison.pdf}
\includegraphics[scale=0.7]{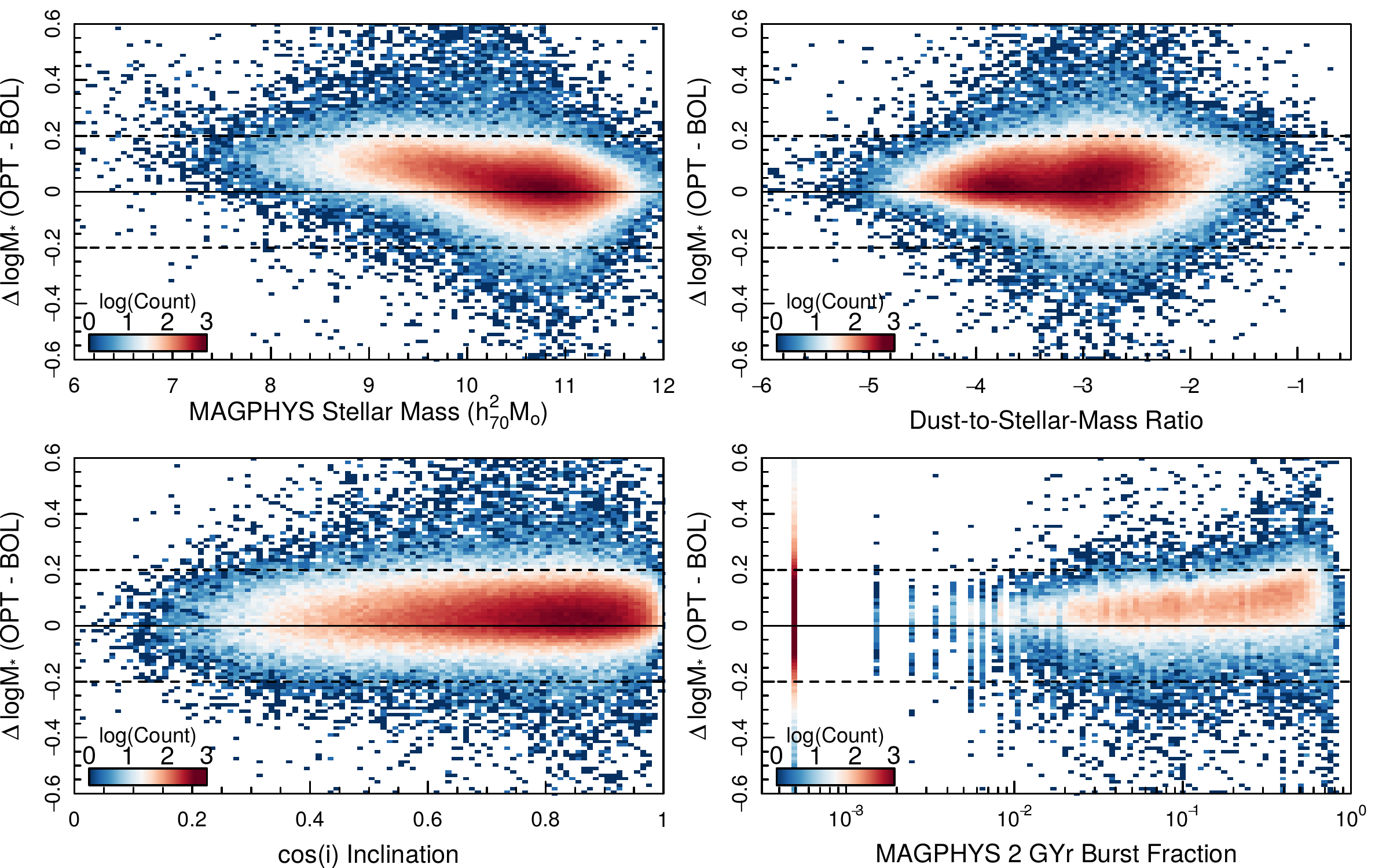}
  \caption[Comparison between
stellar mass estimates in GAMA]{Comparison of bolometric stellar masses returned by
\protect \magphys\ to those measured using the optical-only method presented in
\protect\cite{Taylor2011}, as a function of \magphys\ stellar mass (panel `a'),
\magphys\ dust mass (panel `b'), galaxy inclination (panel `c'), and \magphys\
burst fraction over the last 2 Gyr (panel `d'). While all of these figures show typical agreement within
$\pm 0.2$ dex, there are systematic trends visible in each distribution which we
attribute to the {\bbf difference in chosen dust attenuation and burst models} between the codes.
}\label{fig: stellarmasses} \end{figure*}

Figure \ref{fig: stellarmasses} shows a compendium of the four
main comparison planes that demonstrate systematic differences; namely
variations as a function of stellar mass (upper left), dust-to-stellar mass ratio (upper right), galaxy
inclination (lower left), and {\bbf \magphys\ burst fraction over the last 2 Gyr} (lower right).
We note that there are two populations that separate
out in the upper panels, most notably in the dust-to-stellar mass ratio comparison.
The most systematically different stellar masses are localised at small stellar
masses, high dust-to-stellar mass ratios, and {\bbf at higher \magphys\ burst fraction}. Each of these
properties is consistent with belonging to the predominantly young and
disc-dominated portion of the sample, where bursts and variations in the dust
obscuration prescription are likely to have the most impact. As a result, we
postulate that the differences seen in the mass estimates stem predominantly
from the differences in {\bbf libraries, models, and burst prescriptions } implemented in our
fitting procedures. However
despite these visible differences, we find that $94.8\%$ of the sample are
contained within $\lvert{\Delta \log_{10} M}\rvert \le 0.2$ for the entire
sample. This fraction increases to $97.8\%$ if we select only masses with
\magphys\ goodness-of-fit $0.5 \le \chi_\nu^2 \le 1.5$. For the low redshift
portion of the data ($0.002<z<0.1$), there are $86.0\%$ of masses within $\lvert{\Delta\log_{10}  M}\rvert \le
0.2$, and $88.8\%$ when selecting $0.5 \le \chi_\nu^2 \le 1.5$.  In general, the
optically derived masses return slightly higher stellar masses
(median offset $\Delta\log_{10} M = \log_{10} M_{\rm OPT} - M_{\rm BOL} = 0.03$)
than the bolometrically modelled masses, and (as there is no obvious trend in inclination) there appears
to be no indication of significant quantities of optically thick dust. All of these systematic
shifts in masses are well within both the typical quoted mass uncertainty (median mass uncertainty $ \delta \log_{10} M = 0.10$),
and within the width of the central $68^{\rm th}$ percent range (\ie $1\sigma$) of the distribution ($\sigma_M = 0.14$).

Finally, we implement a correction to account for flux/mass missed by the matched
aperture photometry described in \cite{Wright2016}. To correct for systematically
missed flux/mass, we utilise the GAMA Sersic profile fits to our sample. We calculate
the linear ratio between the measured Sersic flux and aperture flux for each source
(this is the same aperture correction described in \citealt{Taylor2011}, and is often
referred to as the `fluxscale' factor in GAMA data products and publications). This correction
has the effect of preferentially boosting high-mass sources, as stellar mass is loosely
correlated with galaxy Sersic index {\it n} and (in a fixed finite aperture) galaxies will
increasingly miss flux with increasing {\it n}. However, as this correction is based on the
empirically estimated Sersic fits (which are themselves possibly subjected to random and systematic
biases), we provide the results for the uncorrected masses in Appendix \ref{appendix: nofluxcorr}.
These fits provide lower limits for the various parameters estimated in this work.

\subsection{Additional systematic biases} %{{{
{\bbf
By estimating our stellar masses using our `optical' and `bolometric' methods, we attempt to explore
how the stellar mass function is affected by some of the choices and assumptions that have been
made in this work (such as the impact of dust and the allowed burstiness).
However these tests certainly do not encompass the full gambit
of assumptions implicit to stellar mass estimation using stellar population synthesis (SPS) models.
Such assumptions are required
because of our uncertainty of, for example,
the stellar initial mass function \citep{Driver2012}, the contribution of thermally-pulsing asymptotic
giant branch (TP-AGB) stars \citep{Maraston2005,Bruzual2007,Conroy2009}, the choice of
parametrization of star formation histories \citep{Fontana2004,Pacifici2015},
modelling of bursts \citep{Pozzetti2007}, and more. Here we briefly discuss the
effect of some of these assumptions, and derive an estimate of the systematic uncertainty required to be
added to our estimates of stellar masses and their derived quantities.

Systematic effects originating from our uncertainty in the stellar IMF are well documented in the
literature, and there is an ongoing debate as to whether the shape of the initial mass function is
well described by something akin to the \cite{Chabrier2003} IMF, or whether it is better described by a
top-heavy \citep{Baldry2003} or bottom-heavy \citep{Kroupa1993} function, or whether there is a
single valid description for the IMF over all times \citep{Wilkins2008}.
Generally, variation of the IMF manifests itself
as a shift in the stellar population mass-to-light ratio, and thus as a scaling of the estimated mass of each
galaxy, as the IMFs typically differ in their treatment of only the most and least massive stars
\citep{Bell2003,Driver2013}. This, in turn, means that a change in the IMF will cause a multiplicative scaling
of estimated quantities such as {$\mathcal M_{\star}$} and {$\Omega_\star$}. \cite{Driver2012} provided a
prescription for converting between some of the various popular IMFs in the literature, which we reproduce
here in Table \ref{tab: IMFs}. By providing this table we wish to emphasise that the estimates of
{$\mathcal M_{\star}$} and {$\Omega_\star$} provided in this work are valid only for the
\cite{Chabrier2003} IMF, and that these values are highly sensitive to the choice of IMF. Nonetheless,
in the case where the variation in the IMF can be well described by a multiplicative scaling of
overall stellar mass, Table \ref{tab: IMFs} should allow the conversion of our estimated parameters
between the \cite{Chabrier2003} and other popular IMFs. Note that as these corrections
are only valid in the case of a single non-evolving IMF, or when analysing galaxies over a fixed epoch
\citep{Conroy2009}.
}

\begin{table}
\centering
\begin{tabular}{c|c}
IMF & $M_{\star}/M_{\star,Ch}$ \\
\hline
\cite{Salpeter1955} & 1.53 \\
\cite{Kroupa1993} & 2.0 \\
\cite{Kroupa2001} & 1.0 \\
\cite{Chabrier2003} & 1.0 \\
\cite{Baldry2003} & 0.82 \\
\cite{Hopkins2006} & 1.18 \\
\hline
\end{tabular}
\caption{Multiplicative factors for converting between stellar masses and mass densities that are
estimated using different initial mass functions,
relative to the \protect\cite{Chabrier2003} IMF used in this work. }\label{tab: IMFs}
\end{table}

{\bbf
In addition to the uncertainty about the shape of the IMF, additional SPS uncertainties can lead to
significant systematic biases in stellar mass estimation. \cite{Conroy2009} provide a detailed discussion
of uncertainties in SPS masses related to, in particular, TP-AGB stars, horizontal branch stars, and blue
straggler stars. Each of these populations are poorly constrained in SPS models, due to their rarity and
difficulty to constrain observationally. \cite{Conroy2009} conclude that the typical uncertainty
on their mass estimates at $z\sim0$ range from $\sim 0.1$\ dex to $\sim0.4$\ dex, at $95\%$ confidence and for
a range of galaxy colours and magnitudes, due to uncertainty in each
of these parameters. Furthermore, this uncertainty is not restricted to stellar masses estimated using
SPS models. \cite{Gallazzi2009} show that, using either spectral or photometric estimates of stellar
mass-to-light ratios,
one can reach a limiting accuracy of only $\sim0.15$\ dex in the regime where a galaxy has undergone recent
bursts of star formation. Galaxies with more passive histories can be more accurately constrained
in their study; the highest signal-to-noise sources which, are dominated by an old stellar population, having
constraints better than 0.05\ dex (albeit without consideration of the effects of TP-AGB or HB stars, nor the
impact of dust).

Given these systematic biases in estimating stellar mass, it is therefore necessary to encode our
systematic uncertainty into our results, separate from the uncertainty due to our fitting and sample.
Therefore, throughout this work we will consistently provide two uncertainties on each of our
estimates of {$\mathcal M_{\star}$} and {$\Omega_\star$}; \ie
{$\Omega_\star = {\rm Val} \pm \sigma_{\rm fit} \pm \sigma_{\rm sys}$}. Here $\sigma_{\rm fit}$ is the
uncertainty due to our sample and fitting procedure, and incorporates both random uncertainty due to the
fit optimisation (discussed further in Section \ref{sec: uncs}), and the uncertainty due to cosmic variance
(where relevant). For the
parameter $\sigma_{\rm sys}$ we choose a fairly conservative $0.2$\ dex ($58\%$) uncertainty, encompassing
those expected by both \cite{Conroy2009} and \cite{Gallazzi2009}. This value is large, easily dominating over
uncertainties quoted on {$\mathcal M_{\star}$} and {$\Omega_\star$} in previous works that did not incorporate
a quantification of this uncertainty. This is an indication that the uncertainty on our measurement is likely
to be dominated by these systematics, and that improvement in the estimation of {$\Omega_\star$} in particular
will be limited by the reduction of uncertainty of stellar mass modelling in the future.

Finally, we also quantify the systematic uncertainty on the Schechter function slope parameters $\alpha_1$ and
$\alpha_2$ (see Section \ref{sec: schechter func}).
While a constant systematic bias in stellar mass will not cause a change in the Schechter function slope, a
mass-dependent systematic bias may have this effect. To quantify the slope systematic
uncertainty we measure the change in mass function slope when applying a mass dependent systematic bias
of the form $M_{\star,\rm sys} = 0.95\times M_\star + C_1$ and $M_{\star,\rm sys} = 1.05\times M_\star + C_2$, where
the constants $C_1$ and $C_2$ are chosen such that $\langle M_{\star,\rm sys} \rangle = \langle M_{\star} \rangle$.
These functions bias our stellar masses by $\Delta M_\star > 0.2$ dex across the mass range of
our sample, and so simulate a mass dependent systematic bias on the same scale as the (conservative)
systematic mass bias we adopted for {$\mathcal M_{\star}$} and {$\Omega_\star$}. Fits to these biased
masses do exhibit a change in the Schechter function slope parameters, and indicate a conservative systematic
uncertainty on $\alpha$ is $\sigma_{\rm sys} = 0.15$. We adopt this value for the remainder of this
work.

}
%}}}
%}}} }}}
\section{The Density-Corrected Maximum-Volume GSMF}\label{sec: Two} %{{{
%What is the method? {{{
Our primary method to calculate the GSMF uses a density-corrected maximum-volume
(DCMV) weighting to determine the number density distribution of sources,
corrected for absolute-magnitude based observational biases (\ie
\cite{Malmquist1922} bias). The typical maximum-volume corrected number density
\citep{Schmidt1968} is calculated by weighting each galaxy by the inverse of the
comoving volume over which the galaxy would be visible, given the magnitude
limit of the sample, $1/V_{\rm max,i}$. \cite{Saunders1990} and \cite{Cole2011} extend this
method to correct for the presence of over- and under-densities in the radial
density distribution caused by large-scale structure. This is done by defining a
fiducial density between two redshift limits $z_a$ and $z_b$, and using the
ratio of instantaneous density to fiducial density to weight sources, thus
avoiding bias due to over- and under-densities caused by large-scale structure.
\cite{Weigel2016} showed that this method is robust to observational biases, and
indeed returns fits equivalent to those returned by more complex methods, such
as the stepwise maximum likelihood method described by \cite{Efstathiou1988}.

\begin{figure}
\centering
\includegraphics[scale=0.17]{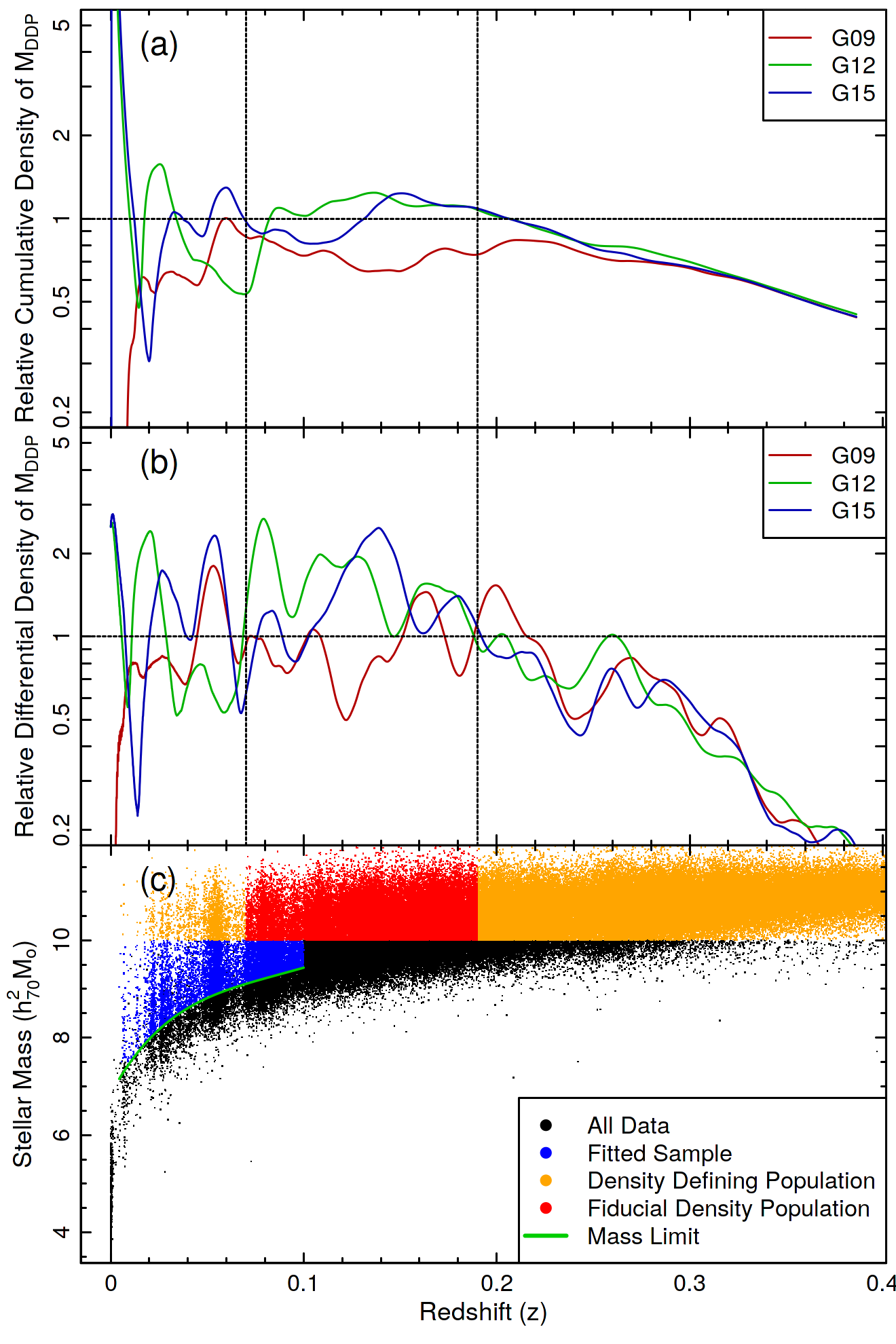}
\caption[Running cumulative and differential density of the GAMA low-z
fields]{The running density of the GAMA data in each of the 3 equatorial GAMA
fields. Panel (a) shows the individual cumulative densities of each field
separately, relative to the fiducial density. Panel (b) shows the relative
differential density of each field.  In the first two panels, the dotted
vertical lines mark the redshift boundaries of our fiducial sample. Panel (c)
shows the distribution of stellar mass against redshift, with the sample used
for estimating the fiducial density highlighted in red, the density defining population
in orange, and our low-z mass-limited sample
highlighted in blue. The green line shows our mass-limit function used in
fitting the GSMF.  From these distributions, we conclude that the fiducial
sample is not adversely impacted by substantial stellar mass incompleteness or
variations in density.}\label{fig: squiggle} \end{figure}

%}}} Define the GSMF and the weights used {{{
The DCMV GSMF is defined by first calculating an individual weight for each
source in our sample. The DCMV weight per object is
\begin{equation}\label{eqn:Vprime}
W_i = \left(V^\prime_{\rm max}\right)^{-1} = \left[\frac{1}{V_{\rm max}}
\frac{\langle\delta_f\rangle}{\delta_i}\right],
\end{equation} where
$V_{\rm max}$ is the standard maximum-volume factor from \cite{Schmidt1968},
$\delta_i$ is the instantaneous running density of galaxies at the redshift of
galaxy $i$, and $\langle\delta_f\rangle$ is the average density of a chosen
fiducial population.  In this work, we define this fiducial average density
$\langle\delta_f\rangle$ using the sample of GAMA targets with $M_\star >
10^{10} M_\odot$ and $0.07 < z < 0.19$. We choose this sample because it
exhibits a fairly uniform density, is not affected by incompleteness, and is
affected by cosmic variance at the $< 10\%$ level (using the cosmic variance
estimator of \citealt{Driver2010}; accessible at \url{cosmocalc.icrar.org}).
{\bbf Nonetheless, cosmic-variance remains a non-negligible source of uncertainty and
therefore is incorporated into all relevant parameter estimates. }
Panel (a) of Figure \ref{fig: squiggle} shows the relative cumulative density of
each of the 3 GAMA equatorial fields, and the region over which our fiducial
density is determined. Similarly, Panel (b) shows the differential running
density of each field. Finally, panel (c) shows the fiducial sample in
mass-redshift space, and shows that the sample is complete in this redshift
range.

The cumulative density distributions of each GAMA equatorial field indicate that,
integrating the number density out to $z=0.1$,
G12 is over-dense relative to our fiducial density by
a factor of $1.02$, while G09 and G15 are under-dense relative to our fiducial density by a factor
of $1.36$ and $1.22$ respectively. This inter-field variation is in good agreement
with the expected cosmic variance between the GAMA fields, which is $\sim23\%$ per field
using the cosmic variance estimator of \cite{Driver2010}.

%}}} Fitting Process: ML {{{
\subsection{Schechter function formalism} \label{sec: schechter func}
In this work we will fit mass
functions to a range of samples. For this, we {\bbf elect to use a 2-component }
\cite{Schechter1976} function. The Schechter function is a specialised form of
the logarithmic truncated generalised gamma distribution (TGGD\footnote{
R package {\tt TGGD} is available on the Comprehensive R Archive Network (CRAN)};
\citealt{Murray2016}):
\begin{equation}\label{eqn: tggd}
\Gamma^t(x;\alpha,\beta,s,m) =
\frac{\log\left(10\right)\beta\left(10^{\left(x-s\right)}\right)^{\alpha+1}
\exp\left(-10^{\left(\beta\left(x-s\right)\right)}\right)}{s\Gamma\left(\frac{\alpha+1}{\beta},\left(10^{(m-s)}\right)^\beta\right)},
\end{equation} where $\Gamma$ is the incomplete upper gamma distribution,
$\alpha$ is the power-law slope of the TGGD, $\beta$ is the rate of exponential
cut-off of the TGGD, $s$ is the scale factor that determines the transition
point between the power-law and exponential regimes, and $m$ is the lower-limit
that defines the truncation point of the TGGD.  {\bbf The TGGD reduces to the standard
Schechter function }
when $\beta=1$, and (in this form) the TGGD parameters
$\alpha$ and $s$ reduce to the normal Schechter parameters $\alpha$ and
${\mathcal M}^\star$, and we define $m$ to be the minimum mass used in our
sample ${M}_{\star}^{\rm min}$ ;
\begin{equation}\label{eqn: Schechter}
S(M;{\mathcal M}^\star,\alpha,M_{\star}^{\rm min}) \sim \Gamma^t\left(M;\alpha,1,{\mathcal M}^\star,M_{\star}^{\rm min}\right).
\end{equation}
As we are using the logarithmic TGGD, masses $M$, ${\mathcal M}^\star$, and
$M_\star^{\rm min}$ are all assumed to be logarithmic also.
{\bbf We choose to formulate the Schechter function in this way (\ie described
using the specialised form of the
logarithmic TGGD, rather than using a directly defined Schechter function)
as the TGGD is a fully analytic PDF, where the normalisation parameter is able to
be evaluated at arbitrary $\alpha$ and ${\mathcal M}^\star$ as: }
\begin{equation}\label{eqn:phistar}
\phi^\star = \frac{\Gamma^t\left({\mathcal M}^\star;\alpha,1,{\mathcal
M}^\star,M_{\star}^{\rm min}\right)}{\log\left(10\right)\exp\left(-1\right)}.
\end{equation} {\bbf Thus this formulation does not require any (often CPU intensive) numerical
integration to estimate the function normalisation. }
Using the TGGD to describe the single Schechter function, we
define the double Schechter as the sum of two single Schechter functions, with
a fractional contribution of component 1,
$f_{\rm mix}$, integrated down to $M_{\star}^{\rm min}$:
\begin{multline}\label{eqn: schecht}
S^d(M;{\mathcal M}^\star,\alpha_1,\alpha_2,f_{\rm mix},{M}_{\star}^{\rm min}) = \\
S\left(M;{\mathcal M}^\star,\alpha_1,{M}_{\star}^{\rm min}\right)\times f_{\rm
mix} + \\
S\left(M;{\mathcal M}^\star,\alpha_2,{M}_{\star}^{\rm
min}\right)\times\left(1-f_{mix}\right).
\end{multline}
The double Schechter
function is useful for fitting distributions that are expected to contain
multiple components, but which we elect to
fit with a coupled ${\mathcal M}^\star$. This has become somewhat common practice in the
literature (see, \eg, \cite{Peng2010b,Baldry2012,Eckert2016}), and we follow this procedure as it enables us
to more readily compare our results with these previous GSMF estimates.{\bbf
Nonetheless, fits with a decoupled ${\mathcal M}^\star$ have merit, and can encode interesting physics
(see, \eg, \citealt{Kelvin2014,Moffett2016}). We therefore opt to include the decoupled GSMF fits in
Appendix \ref{appendix: decoupled}, for examination by the interested reader. }

{\bbf Our formulated } distribution can then be fit to individual data in two ways: by specifying
individual weights based on some relevant criteria (\eg density corrected
maximum volume weights) and fitting over a fixed mass range, or by defining an
expected limiting stellar mass ${M}_{\rm \star,i}^{\rm lim}$ per source (\eg
where observational incompleteness becomes important for that source, in the
mass plane). In the latter case, the log-likelihood of each source is then
calculated with consideration of the limiting stellar mass of that source {\em
given} the shape of the Schechter function at that iteration. In this way, the
latter procedure includes information of the mass function in the optimisation
process in a more considered fashion than the former {\bbf (the optimisation method is
discussed in Section \ref{sec: optimisation})}. We therefore fit our
distributions using the mass limit optimisation procedure, whereby  we define
limits using an analytic expression similar to that of \cite{Moffett2016}, but
modified to match this sample of masses (see Section \ref{sec: mass limits}).
%We describe the process of generating mass limits in section \ref{sec: masslimits}.
Note, however, that as we no longer have a single fixed mass limit,
our mixture fraction $f_{\rm mix}$ must now be modified per object to reflect
the effective mixture fraction given each individual source's mass limit,
$f_{\rm mix,i}$:
\begin{gather}
I^{\rm mix}_{\rm 1,i} = f_{\rm
mix}\times\int\limits_{\rm M_\star^{\rm min}}^{M_{\rm \star,i}^{\rm
lim}}\!\!\!\! S\left(M;{\mathcal M}^\star,\alpha_1,M_\star^{\rm min}\right){\rm
d}M, \\
I^{\rm mix}_{\rm 2,i} = \left(1-f_{\rm mix}\right)\times\int\limits_{\rm
M_\star^{\rm min}}^{M_{\rm \star,i}^{\rm lim}}\!\!\!\! S\left(M;{\mathcal
M}^\star,\alpha_2,M_\star^{\rm min}\right){\rm d}M,
\\
f_{\rm mix,i} = \frac{I^{\rm mix}_{\rm 1,i}}{I^{\rm mix}_{\rm 1,i} + I^{\rm mix}_{\rm 2,i}}.
\end{gather} Using these individualised limits and mixture fractions, we define
the log-likelihood of our fit:
\begin{equation}\label{eqn: loglike}
{\rm ln \mathcal L} = \displaystyle\sum\limits_i \log\left[S^d\left(M_i;{\mathcal
M}^\star,\alpha_1, \alpha_2,f_{\rm mix,i},M_{\rm \star,i}^{\rm
lim}\right)\right]\times\frac{\delta_i}{\langle\delta_f\rangle},
\end{equation}
and optimise simultaneously for ${\mathcal M}^\star$, $\alpha_1$, $\alpha_2$,
and $f_{\rm mix}$. The primary benefit of implementing the mass limits in this way
is that it at no point requires binning of the data in any form.
After this optimisation, we can calculate the values of
$\phi^\star_1$ and $\phi^\star_2$ using the fit parameters and the defined
fiducial population number density, recognising that the ratio of $\phi^\star$
values is directly proportional to the integral of the individual Schechter
components:
\begin{gather}\label{eqn: phirat} I^{\rm mix}_{\rm 1} = f_{\rm
mix}\times\int\limits_{\rm M_{\star}^{\rm min}}^{\inf}\!\!\!\!
S\left(M;{\mathcal M}^\star,\alpha_1,{M}_{\star}^{\rm min}\right){\rm d}M, \\
I^{\rm mix}_{\rm 2,i} = \left(1-f_{\rm mix}\right)\times\int\limits_{\rm
{M}_{\star}^{\rm min}}^{\inf}\!\!\!\! S\left(M;{\mathcal
M}^\star,\alpha_2,{M}_{\star}^{\rm min}\right){\rm d}M, \\
\frac{\phi^\star_1}{\phi^\star_2}= \frac{I^{\rm mix}_{\rm 1}}{I^{\rm mix}_{\rm
2}}.
\end{gather}

\subsection{Defining Mass Limits}\label{sec: mass limits}%{{{
Our mass limit
function is shown graphically as the green line in panel `c' of Figure \ref{fig:
squiggle} and is known to exhibit $>97\%$ completeness for all sources in GAMA
out to $z = 0.1$, without biases in mass and/or colour. The process for
defining these limits typically involves visually inspecting the distribution of stellar
masses as a function of redshift (and vice versa) and determining the point at
which the sample begins to become incomplete. Once this has been done in a
series of bins of stellar mass and redshift, a polynomial is then fit to the
limits.

However, this process is liable to be biased by the eye of the person estimating the
limits (\ie no two people will be likely to estimate the same limits), and as
such we implement automated methods for determining mass limits. The {\tt
MassFuncFitR} package contains a function that performs the above in an
automated manner, by estimating the turn-over point of the number density
distribution in bins of comoving distance and stellar mass independently.
In each bin of
comoving distance, the function takes the mass at the peak density as the turn
over point, and in bins of stellar mass the function takes the largest comoving
distance at median stellar mass density as the turn-over point. Additionally,
there is the option to bootstrap this estimation procedure to refine the limits.
Indeed, testing of this automated procedure indicates that it is less prone to the
introduction of biases than occurs when fitting for mass limits by hand/eye, {\bbf
and produces a sample that is not biased with respect to colour } (see Appendix \ref{appendix: mass limits}).
%}}}

\subsection{Optimisation procedures} \label{sec: optimisation}
Once we have our per-object weights, we are
able to both visualise and fit the GSMF. For our fits, we utilise a Markov-chain
Monte-Carlo (MCMC).  For our MCMC optimisation, we calculate the best-fit
Schechter function parameters by sampling from the joint posterior-space of the
${\mathcal M}^\star$, $\alpha_1$, $\alpha_2$, and $f_{\rm mix}$ parameters. To
do this we first assign priors to each parameter; we choose to use uniform
priors over the regions $\log_{10} {\mathcal M}^\star \in [8, 11.6]$,
$\alpha_{1,2} \in [-2.5,1.5]$, and $f_{\rm mix} \in [0,1]$.  Given these priors,
evaluated at some sample point, $V_p(\log_{10}{\mathcal M}^\star,\alpha_1,\alpha_2,f_{\rm
mix})$, we can then evaluate the log-posterior as: \begin{equation}\label{eqn:
posterior} {\rm ln}{\mathcal P}= {\rm ln}{V_p} + {\rm ln}{\mathcal L}
\end{equation} where ${\rm ln}{\mathcal L}$ is the same as in Equation \ref{eqn:
loglike}. We sample the posterior space using an Independence Metropolis
sampler, and examine the posterior covariances directly to check for stability.
For our MCMC, we utilise the {\sc Laplace's Demon} package in R, available on
the Comprehensive R Archive Network (CRAN).  Once we have optimised these
parameters, we fit for the total mass function normalisation at ${\mathcal
M}^\star$, and use the $f_{\rm mix}$ parameter to determine the fractional
contributions from each component, thus determining the two $\phi^\star$
parameters.  We then utilise the full posterior distribution to estimate the
uncertainty on each of our $\phi^\star$ values, incorporating consideration for
the covariances between parameters.

%}}}
\subsection{Verifying fit uncertainties}\label{sec: uncs} %{{{
In order to verify that the
uncertainties from our MCMC are a true reflection of the data, we perform 100
Jackknife resamplings of the data, and recalculate our GSMF parameters on the
reduced dataset. The final parameter uncertainties are then compared to the
absolute range in jackknifed parameters. This resampling and re-fitting allows
us to ensure that the MCMC uncertainties are not underestimated, as can be the
case when the likelihood used is not an appropriate reflection of the dynamic
range of the variables being tested, or when the model is not a true generative
distribution for the data. The latter is particularly relevant given that previous
studies indicate that the GSMF is (at simplest) a summation of {\rm many} single component
Schechter functions, rather than just two \citep{Moffett2016}. Therefore, should our
two component approximation be overly-simplistic we may artificially under-estimate
the uncertainties on each of the function parameters.

Furthermore, in our MCMC fits to the double Schechter function we do not
constrain the value of $\phi^\star_1$ or $\phi^\star_2$ directly. Rather, we fit
for the mixture and calculate the normalisations {\em post-facto}. As a result,
we do not directly measure an uncertainty on these parameters either. We
therefore calculate the uncertainties associated with each $\phi^\star$
parameter by calculating the fit (and subsequently the individual component)
normalisations over a range of the possible fit parameters. To do this, we
calculate the normalisation of the fit components for 1000 randomly selected
stationary samples\footnote{\bbf Stationary samples are samples of
the MCMC chains that are deemed to
originate, in the correct proportion, from the true posterior
distribution (the `stationary' distribution).} of the MCMC chains, and use the standard deviation of the fit
normalisations to be representative of the normalisation uncertainty. This
method incorporates all possible covariances between parameters.

%}}}
\subsection{Results of GSMF fits} %{{{
The GSMFs measured using this weighting
method, for the two stellar mass estimation methods, are shown in Figure \ref{fig: gsmf}.
In the figure, we can see that our data are modelled well by the two component
Schechter function, {\bbf and that our two samples are in good agreement regarding their various
fit parameters.}
The best-fit GSMF Schechter parameters for each sample are given in
Table \ref{tab: results1}, along with {\bbf both random and systematic} uncertainties on each parameter, and a
sample of literature GSMF fits, for reference. {Note that the uncertainties in Figure \ref{fig: gsmf} show only the random
component from the optimisation and data/cosmic variance}. We note that our estimate of ${\mathcal M}^\star$
is in tension with some of the previous estimates, being larger than some (\ie \citealt{Baldry2012,Peng2010b})
smaller than others (\ie \citealt{Eckert2016}), and in agreement with the most recent work
from SDSS \citep{Weigel2016}. However, comparison between the quoted uncertainties
of ${\mathcal M}^\star$ from each work with the observed scatter in the estimates themselves suggests that this
tension is likely driven by unquoted systematic uncertainties rather than random uncertainties. Indeed, all
of the quoted ${\mathcal M}^\star$ values agree within our nominal systematic uncertainty of 0.2 dex.

However, one would naively expect the measurements between our dataset and that of \cite{Baldry2012} to be in
reasonable agreement. This is not true with respect to ${\mathcal M}^\star$ in particular.
We argue that this difference is primarily the result of the dedicated by-hand effort which has since been undertaken
to ensure photometry of the brightest systems in GAMA are accurately determined
(see \citealt{Wright2016}). These systems were disproportionately shredded (compared to fainter, smaller systems)
in the original GAMA aperture catalogues. As a result their fluxes were underestimated, and so too their stellar masses.

%Looking at whether we can achieve internal consistency within our {\bbf bolometric and optical mass samples,
%we explore possible explanations for tension between the mass function fits. }
%Examining the joint and marginalised posteriors of the GSMF fit to our
%bolometric masses (Figure \ref{fig: posteriors}), we can see substantial covariance
%between many parameters, as well as some apparent non-gaussianity in the chains. This suggests that there is
%degeneracies in the parameter-space during optimisation, and that the fits are
%being influenced by noise due to our use of uninformative priors.
%We therefore explore whether we are able
%to break this degeneracy, and determine a best-estimate value of the double
%Schechter function parameters.

\begin{figure*} \centering
%\vspace{3in}
%\includegraphics[scale=0.9]{GSMF_fits.pdf}
\includegraphics[scale=0.9]{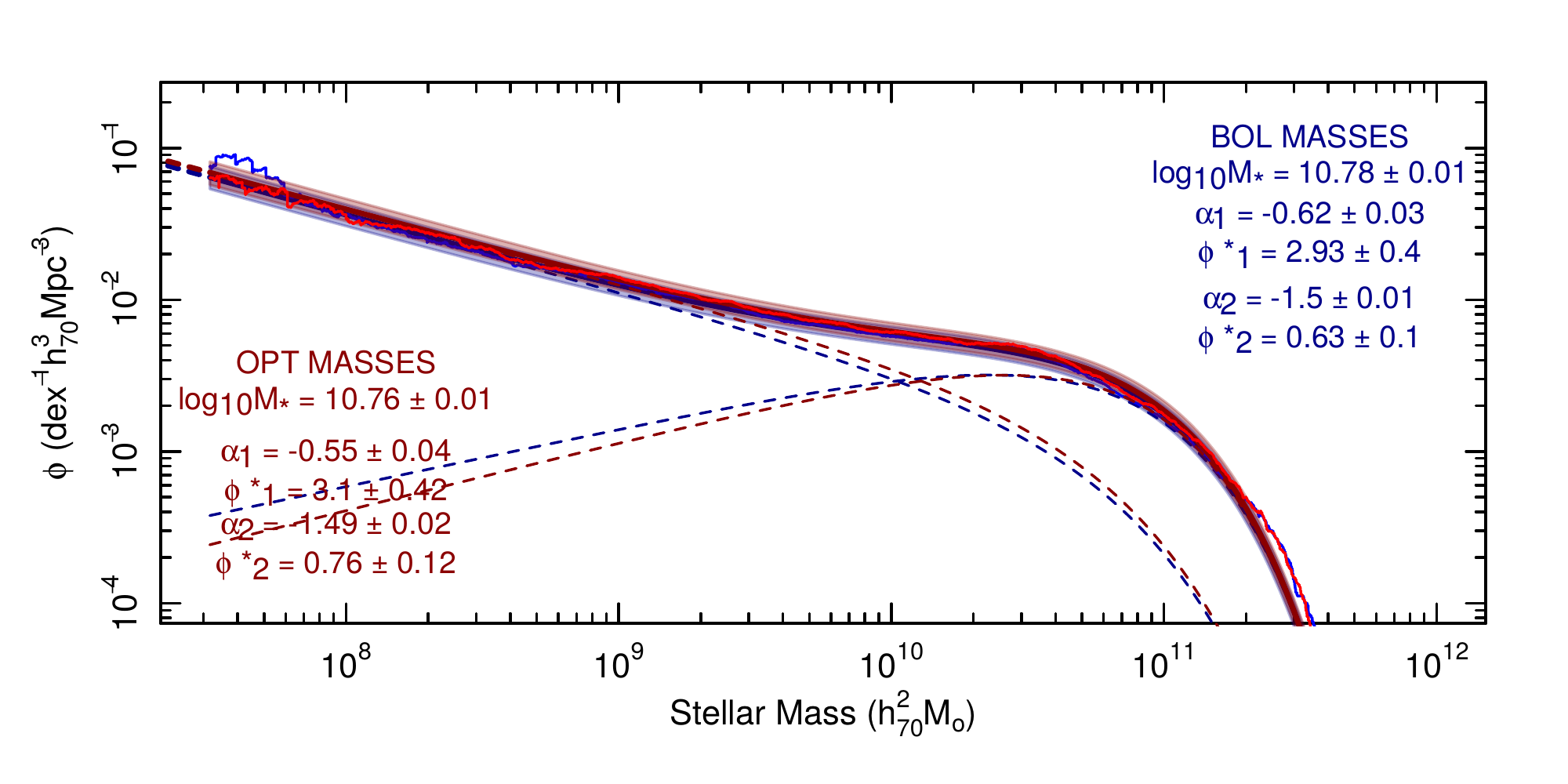} \caption[GAMA low-z stellar mass
functions]{The galaxy stellar mass function fits as estimated using our two mass
samples. The mass function using bolometric \magphys\ stellar masses is shown in blue, and using
our optical masses is shown in red.  Figures are annotated with the fit
parameters and uncertainties, {\em without} inclusion of systematic uncertainties;
\ie these fits {\bbf show only random uncertainties due to fitting and
cosmic variance (as determined using the estimator from \protect \cite{Driver2010}). Our standard
systematic uncertainties on ${\mathcal M}^\star$ and $\alpha$ are not shown.}}\label{fig:
gsmf} \end{figure*}
\begin{table*}
\begin{tabular}{ccccccc}
Dataset & ${\mathcal M}^\star$ & $\alpha_1$ & $\phi^\star_1$ & $\alpha_2$ & $\phi^\star_2$ \\
        &$\left[\log_{10}(M_\odot)\right]$&            &$(\times 10^{-3})$&            &$(\times 10^{-3})$\\
\hline
{{\sc \bf Bolometric}}  & $\mathbf{10.78\pm0.01\pm0.20}$ & $\mathbf{-0.62\pm0.03\pm0.15}$ & $\mathbf{2.93\pm0.40}$ & $\mathbf{-1.50\pm0.01\pm0.15}$ & $\mathbf{0.63\pm0.10}$   \\
{\sc Optical }   & $10.76\pm0.01\pm0.20$ & $-0.55\pm0.04\pm0.15$ & $3.10\pm0.42$ & $-1.49\pm0.02\pm0.15$ & $0.75\pm0.12$ \\
\hline
\cite{Peng2010b} & $10.67 \pm 0.01\pm0.2$ & $-0.52 \pm 0.04 \pm 0.15$ & $4.03 \pm 0.12$ & $-1.56 \pm 0.12 \pm 0.15$ & $0.66 \pm 0.09$ \\
\cite{Baldry2012}& $10.66 \pm 0.05\pm0.2$ & $-0.35 \pm 0.18 \pm 0.15$ & $3.96 \pm 0.34$ & $-1.47 \pm 0.05 \pm 0.15$ & $0.79 \pm 0.23$ \\
\cite{Weigel2016}& $10.79 \pm 0.01\pm0.2$ & $-0.79 \pm 0.04 \pm 0.15$ & $3.35 \pm 2.31$ & $-1.69 \pm 0.05 \pm 0.15$ & $0.17 \pm 0.01$ \\
\cite{Eckert2016}& $10.87^{+0.33}_{-0.27}\pm0.2$ & $-0.52^{+0.87}_{-0.49}\pm0.15$ & $9.00^{+6.36}_{-8.47}$ & $-1.38^{+0.13}_{-0.35}\pm0.15$ & $3.25^{+3.00}_{-2.81}$ \\
\hline
\end{tabular}
\caption[Best-fit Schechter function parameters for the GSMF]{Best fit
parameters of the double Schechter function for our two data sets and fitting
methods, when using density-corrected maximum-volume weights. As a guide, we
also show the double Schechter function fits from \protect\cite{Peng2010b},
\protect\cite{Baldry2012}, \protect\cite{Weigel2016}, and
\protect\cite{Eckert2016}. Our best-fit Schechter function parameters are shown
in {\bf bold}. {\bbf Note these fits show both random uncertainties due to fitting and
cosmic variance (as determined using the estimator from \protect \cite{Driver2010}), and our standard
systematic uncertainties on ${\mathcal M}^\star$ and $\alpha$ due to uncertainty in SPS modelling.
}}\label{tab: results1} \end{table*}

\begin{figure} \centering
%\vspace{3in}
%\includegraphics[scale=0.57]{PosteriorSamples.pdf}
\includegraphics[scale=0.57]{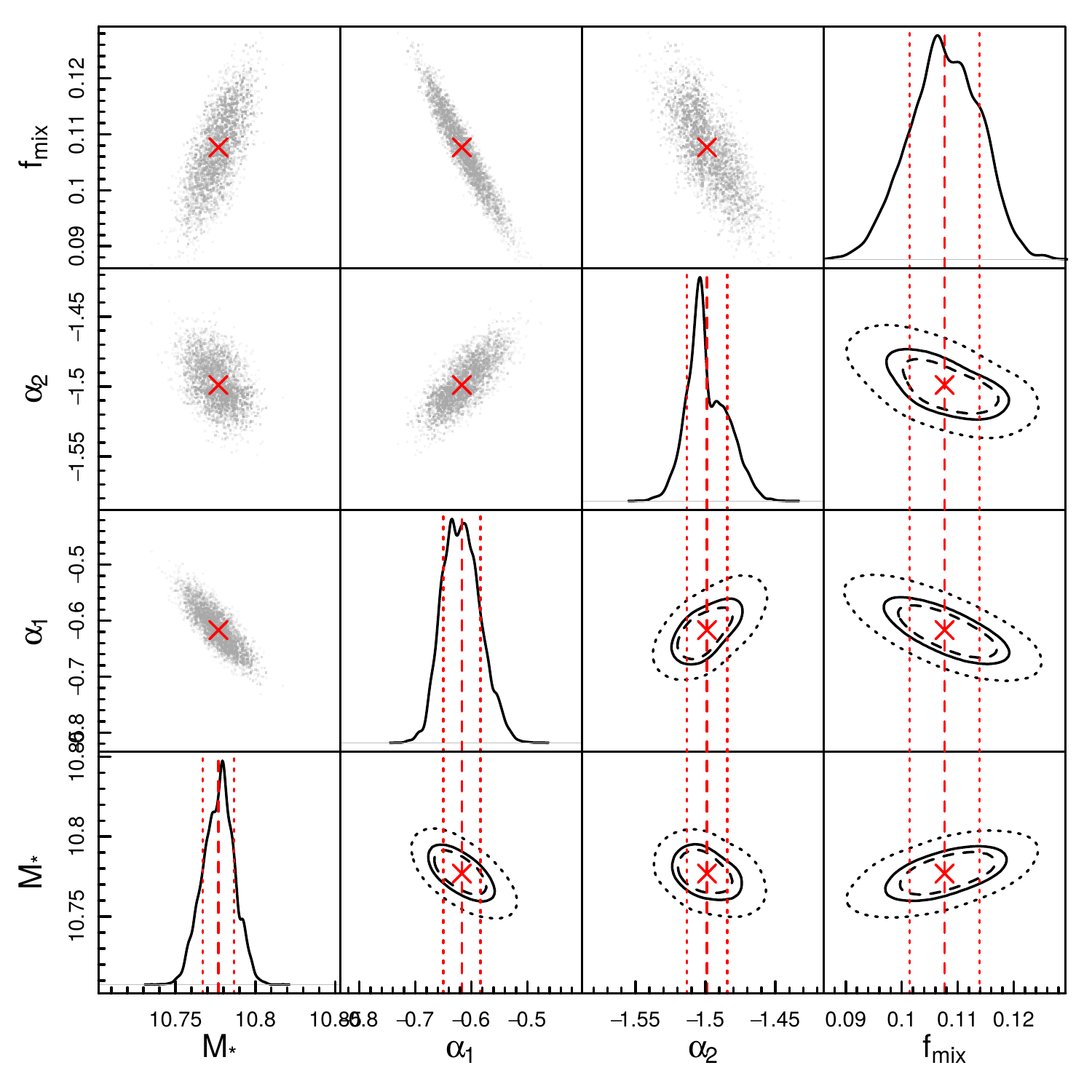} \caption[Posterior samples
from MCMC GSMF fits]{The posterior samples from our MCMC optimisation of the
GSMF using bolometrically estimated stellar masses. Upper triangle: the individual
stationary samples
 (grey points), and the mean of these samples (red cross).
Lower triangle: the contours containing 50, 75, and 90 percent of the posterior samples
(dashed, solid, and dotted lines respectively).
Diagonal: marginalised PDFs of the posterior samples and their mean (red dashed line) and standard deviation (red dotted lines).
}\label{fig: posteriors} \end{figure}
\section{The Volume-Corrected Bivariate Brightness Distribution}\label{sec: Three} %{{{
%What are the methods shortcomings {{{
The DCMV weighting method for estimating the GSMF, as stated in Section
\ref{sec: Two}, incorporates observability corrections based solely on absolute
magnitude. However, we know that there are additional selection effects within
the GAMA sample, specifically around source surface-brightness and compactness.
For example, due to the source definition using SDSS r-band imaging, sources
that have apparent r-band surface brightnesses (averaged within $R_e$) lower
that 23 mag arcsec$^{-2}$ will suffer incompleteness in our sample at the $30\%$
level, and at the $75\%$ level below $24.5$ mag arcsec$^{-2}$
\citep{Blanton2005,Baldry2012,Cross2001}.  In order to investigate these
additional known (and unknown) selection effects into our estimate of the GSMF,
we can derive empirical weights from the data itself and examine the impact this
has on the GSMF.

%}}}
%What is the method? {{{
By plotting the bivariate brightness distribution (BBD) of stellar mass
$M_\star$ and absolute average surface brightness within the effective radius
$\langle \mu_e \rangle_{\rm abs}$, we are able to visualise the majority of the
selection boundaries present in the GAMA data. Panel (a) of Figure \ref{fig:
bbds} shows the observed bivariate brightness distribution for our sample of bolometric
stellar masses, with lines overlaid that mark the selection boundaries of the
sample (see \citealt{Driver1999}). These boundaries are a mixture of
observational unavoidable and intentionally imposed, owing both to the limitations of the
data being analysed and the design of the GAMA survey. However, as these
selection boundaries are typically defined using apparent flux and apparent size
(or variations thereof), the boundaries shown in the absolute
$M-\langle\mu_e\rangle_{\rm abs}$ plane are not sharp; rather they are blurred
systematically as a function of mass-to-light ratio and redshift. We
show the boundaries that would be measured at two characteristic mass-to-light
ratios: $M/L = \{1,3\}$. These mark the {\bbf $\sim90^{\rm th}$ percentile }
limiting $M/L$ values for the GAMA
low-z sample.  From these boundaries we can infer the point of impact of
incompleteness on our sample in $M_\star$-$\langle\mu_e\rangle_{\rm abs}$ space,
and therefore estimate where our analysis becomes biased.  We do this by
examining which selection boundaries intersect with high-density areas of the
BBD. Note also that, while panel (a) suggests that our incompleteness is most
prominent at the spectroscopic and surface-brightness boundaries, to make an
accurate inference we should compare each boundary to the number-density version
of the BBD (\ie panel `c'), rather than the raw-count version, so that we can
see if the post-correction number density is being impinged upon.

%}}} Define the GSMF and weights used {{{
{\bbf In previous studies of the GSMF, estimating surface brightness incompleteness
has sometimes been achieved through simulations. For example, \cite{Blanton2005} do this by assuming a
simple Gaussian analytic form for the surface brightness distribution
of galaxies, and injecting
galaxies sampled from this distribution into their imaging, for extraction and analysis. This
allows an estimation of the fraction of successfully extracted galaxies (as a function of
surface brightness), and thus an incompleteness estimate, to be made.
However, inspection of the measured surface brightness distribution of galaxies shows this
distribution to be somewhat more complex than a simple Gaussian distribution would suggest and that,
indeed, the uncertainty on the true surface brightness distribution means that performing
such an analytic estimate is likely to be biased itself.

Therefore, to estimate our surface-brightness incompleteness, we take a more pragmatic and empirical
approach.} We start by deriving an average weight per bin for each cell in the
BBD.  Within each bin we determine the weighted median redshift, where the
weights are those determined by our density sampling. We then determine the
volume visible to each bin and then divide the summed density-corrected weights
by twice the median volume.  By defining weights in this way, we assume that all
selection effects bias our sample to lower redshift (rather than, \eg cause a
net decrease in number-counts across the entire redshift range) {\bbf and effectively
test the assumption that, in bins of both stellar mass and surface brightness, the distribution
of an unbiased sample of galaxies will have a $V/V_{\rm max}$ distribution that is
uniform over $[0,1]$. If this assumption is correct, then calculating the value of
binned $V_{\rm max}$ in this way should allow us to }
account for all systematic effects in the data, known
or otherwise, without having to explicitly define them. {\bbf In this way our BBD is somewhat
different from a conventionally estimated BBD, such as that presented in \cite{Driver2005}.}
We then use these
weights to calculate the binned number density BBD, and can subsequently
collapse this 2D distribution along the surface brightness axis to recover the
binned stellar mass function. Naturally this is not as statistically elegant as
our first method (in data analysis, not binning is always preferable to
binning), however the exercise is useful in determining if subtle, hidden
selection effects have a substantial impact on the GSMF (compared to just
performing the absolute magnitude based weighting outlined in Section \ref{sec:
Two}).

\begin{figure*} \centering
%\vspace{3in}
%\includegraphics[scale=0.75]{BBDs.pdf}
\includegraphics[scale=0.75]{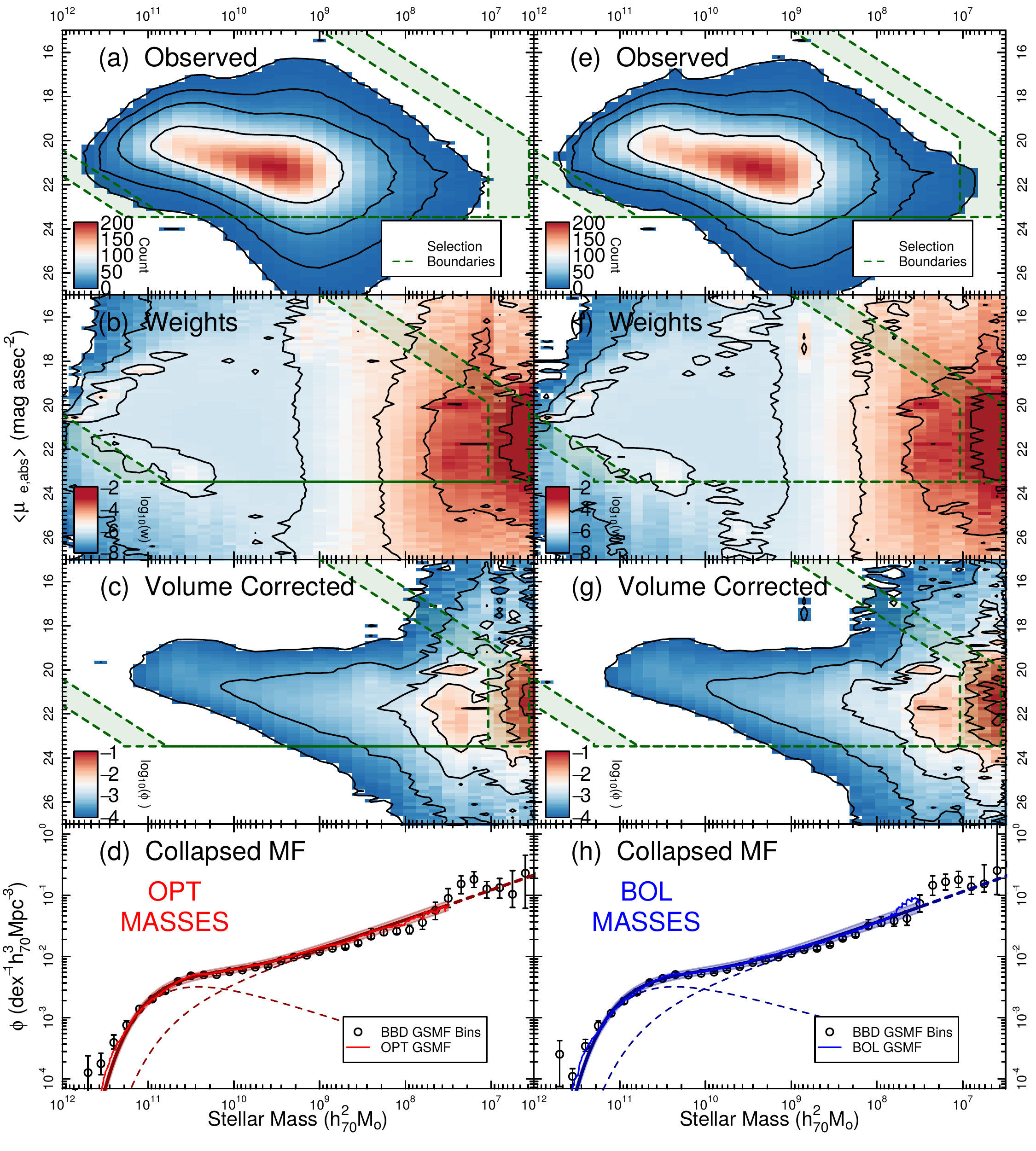}
\caption[The bivariate brightness
distribution of galaxies]{The galaxy bivariate brightness distribution space, {averaged
over MC iterations}, as
a function of raw counts (panels a+e), density-corrected weight per bin (panels
b+f), and number density (panels c+g). Panels d+h show the binned GSMF
determined by collapsing panels c+g onto their respective x-axis.  Overplotted
in these panels is the DCMV GSMF for the same sample, demonstrating agreement
between the mass functions returned using these methods. Note that the
binned GSMFs in panels c+d do not include the additional cosmic variance uncertainty in their
error bars.
}\label{fig: bbds}
\end{figure*}

%}}} Results {{{
Panel (b) of Figure \ref{fig: bbds} shows the weights derived for each bin, and
panel (c) shows the final corrected BBD for the sample. Firstly, we note that
the distribution of weights is not curved or diagonal, but rather exhibits a
fairly linear increase in weight solely as a function of stellar mass. This
suggests that our sample is not strongly sensitive to surface-brightness
effects, even down to our spectroscopic completeness selection limit. {\bbf
Indeed, examination of the distribution of $V/V_{\rm max}$ values in bins of
stellar mass shows a strong evolution, whereas in bins of surface brightness
only a minor change seen. } Secondly,
the number-density distribution in panel (c) appears to be reasonably well
bounded by the two diagonal selection boundaries, as the number density is
declining well before these limits. This suggests that there is not a
substantial population of massive low surface brightness galaxies, nor highly
compact galaxies, that we have missed because of selection effects. Naturally
this does not exclude that these galaxies can exist (indeed, rare examples of
extremely massive low surface brightness galaxies have been known to exist for
decades; see \citealt{Bothun1987}) but rather suggests that they do not
contribute greatly to the number-density of galaxies
\citep{Cross2001,Driver1999,Davies2016b}.  Finally, panel (d) of Figure
\ref{fig: bbds} shows the binned-GSMF measured from the BBD using our bolometric
masses. This is shown jointly with our DCMV GSMF, as a demonstration of the
agreement between these analysis methods in the $M_\star > 10^{8} M_\odot$
regime.  {\bbf There is a slight indication of a possible excess in the BBD GSMF at
masses below $10^{8} M_\odot$, suggesting that incompleteness may likely be affecting our sample
below this point. }

Panels (e)-(h) of Figure \ref{fig: bbds} show the same as (a)-(d), but
for our optical-based sample of stellar masses. For this sample we can see the same trends
as for the bolometric  mass sample, and similarly good agreement between the two GSMFs
for this sample.

%}}}
%%}}}
\subsection{Extension to Future Surveys}\label{sec: discussion}%{{{
We have
derived four estimates of the GSMF for the full GAMA $0.002 < z < 0.1$ sample,
summarised in Table \ref{tab: results1}.  We find GSMFs that are in agreement
with previous GAMA estimates, for fits to a limiting mass of $10^{7.5} M_\odot$.
{\bbf However, there is a suggestion that we may be incomplete below $10^{8} M_\odot$,
where we become restricted in
our fitting power by the surface-brightness limit of SDSS imaging (which was
used to select the GAMA sample)}. %While
%our empirical checks in Section \ref{sec: Three} suggests that this is not a
%major factor, we cannot rule out some subtle bias.

\begin{figure*} \centering
%\vspace{3in}
%\includegraphics[scale=0.9]{FutureSurveys.pdf}
\includegraphics[scale=0.9]{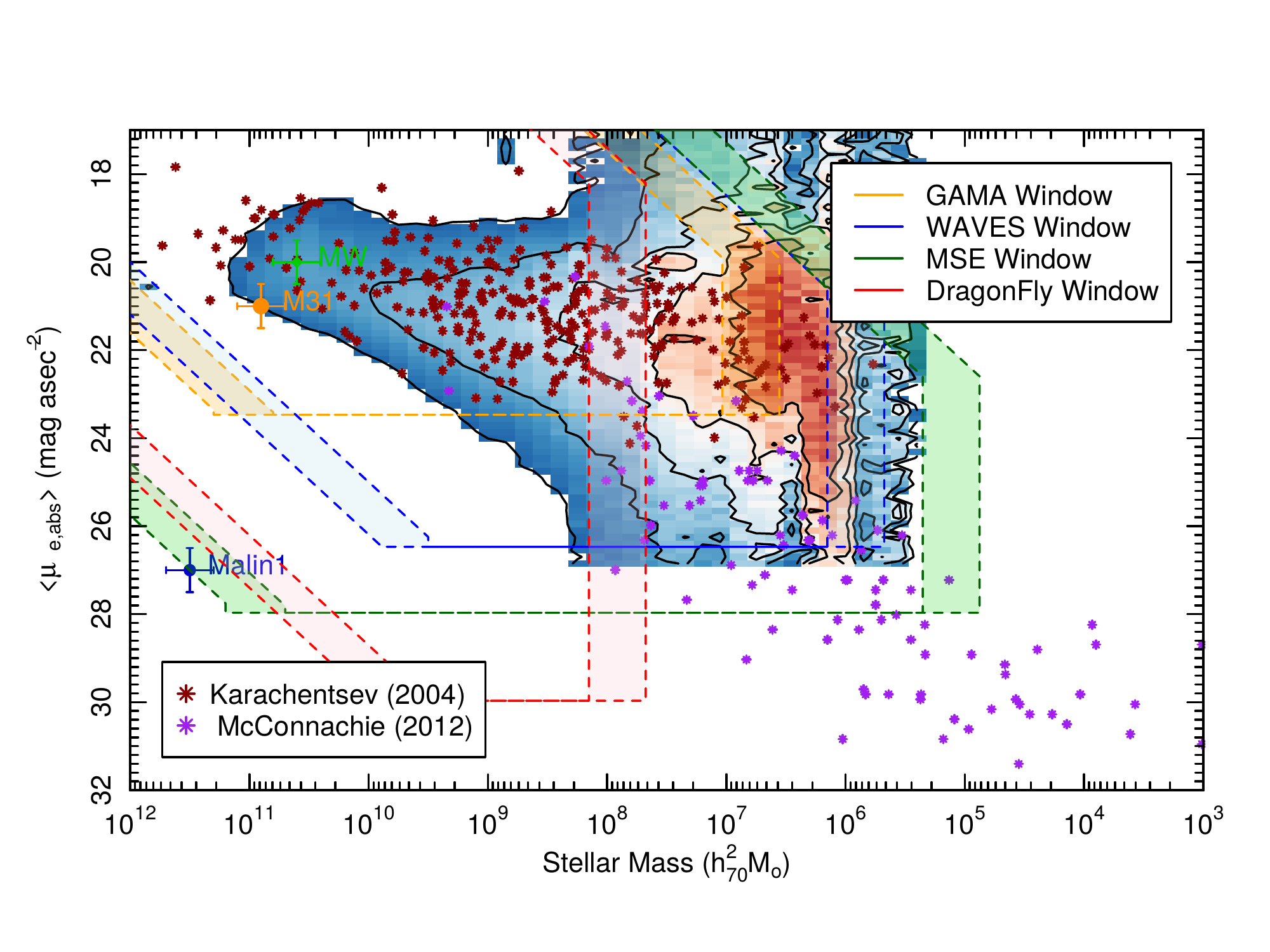}
\caption[The GAMA BBD with future
survey parameter-space]{The bivariate-brightness distribution for the GAMA $z <
0.1$ sample. We overlay the selection boundaries for GAMA, as well as the
expected limits for three future surveys: the WAVES-Wide survey on 4-MOST, a
similar-time survey on MSE, and the Dragonfly LSB survey.  Furthermore, we
overplot data from the local sphere and local group as a demonstration of where
as-yet-undetected galaxies beyond the local sphere are expected to lie, as well
as some individual galaxies of note: Milky Way (green dot), Andromeda (orange
dot), and Malin 1 (blue dot).}\label{fig: selections} \end{figure*}

Our estimates of the GSMF are predominantly limited by the selection boundaries
in spectroscopic completeness and surface brightness. Figures \ref{fig: bbds}
and \ref{fig: selections} demonstrate these selection boundaries, as well as
other boundaries that affect our analysis to a lesser degree (namely
compactness, sparseness, and rarity).  Despite these limits, however, we are able to
construct a GSMF that is representative down to masses as low as $\sim 10^{6}
M_\odot$, by simply continuing our empirical reconstruction beyond $10^{7.5}
M_\odot$, as we will show in Section \ref{sec: g10}. There is evidence of a
systematic incompleteness bias below $10^{8} M_\odot$, which confirms our
concerns regarding incompleteness, but nonetheless the mass function shows
continuity consistent with the extrapolation below this limit (\ie the impact is
subtle, not severe).

Because of the incompleteness effects in GAMA, it is desirable to extend this
work using future deep large-area surveys if we wish to constrain the GSMF to
yet lower masses using a single sample. To demonstrate this, Figure \ref{fig:
selections} shows the selection boundaries for two future surveys: the Wide Area
Vista Extragalactic Survey (WAVES; \citealt{Driver2016}), and the galaxy
evolution survey on the Mauna Kea Spectroscopic Explorer (MSE;
\citealt{McConnachie2016}). WAVES and MSE will both utilise imaging that is
substantially deeper than the GAMA SDSS imaging, and will have high-completeness
spectroscopic campaigns that push many magnitudes fainter than was possible for
GAMA. As a result, these surveys will both substantially expand the available
parameter space available to be studied for galaxy evolution, as can be seen by
the expansion of the limits in Figure \ref{fig: selections}. As a demonstration,
we include galaxies measured in the local-sphere in this figure, to indicate
where it is expected that the majority of galaxies might lie in this plane
(beyond the limits of GAMA). For these points, we have used the local sphere
catalogue from \cite{Karachentsev2004} and the ``maintained'' local group sample
from \cite{McConnachie2012}.  Finally, we include the selection-boundaries of a
low-surface brightness survey using the Dragonfly telephoto array
\citep{Abraham2014}, which clearly opens up a very different part of the
parameter space.

{\bbf The samples of \cite{Karachentsev2004} and \cite{McConnachie2012} are
particularly useful in inferring the likely incompleteness of our sample.
In particular, it is telling that half of the \cite{McConnachie2012} sample with mass
greater than $10^{7.5} M_\odot$ lies below our nominal surface brightness completeness
limit. This
provides further suggestion that our sample is likely incomplete below this level.
It is clear that the next generation of wide-area spectroscopic
surveys, such as WAVES-wide, will be paramount in determining the shape of
the low-mass tail of the stellar mass function. Prior to the execution of these
large surveys, however, we can perform a similar analysis by combining the wide-area power of
GAMA with a more directed, deeper survey, such as the G10-COSMOS. }

\begin{table*} \caption{Survey parameters of GAMA, G10-COSMOS, and 3 additional
surveys}\label{tab: Survey designs} \begin{tabular}{c|ccccccc} Survey & Area &
Selected & Spec. Limit & Surf. Brightness & Resolution & Pixel Width &
Completeness \\ & (deg$^2$) & From  & (mag)       & Lim. (mag/arcsec$^2$) &
($^{\prime\prime}$) & ($^{\prime\prime}$) & ($\%$ within limit) \\ \hline GAMA
& 180 & SDSS & 19.8 & 24.5 & 1.2 & 0.339  & $>98$ \\ G10-COSMOS  & 1 & HST &
24.5 & 24.5 & 1.2 & 0.339 & $\sim40$ \\ DragonFly  & 180 & SDSS & 19.8 & 30.5 &
5.5 & 2.3 & $>98$ \\ WAVES-Wide & 1500 & VST KiDS & 22 & 26.5 & 0.6 & 0.2 &
$>95$ \\ MSE & 12000 & LSST & 24 & 28 & 0.6 & 0.2 & $>95$ \\ \hline
\end{tabular} \end{table*}

\subsection{Exploiting GAMA + G10-COSMOS }\label{sec: g10} While we will
require surveys like WAVES and MSE in order to constrain the GSMF in a robust
fashion below $10^{7.5} M_\odot$ using a single dataset, we note that by
splicing our GAMA equatorial sample with the G10-COSMOS sample of
\cite{Andrews2016}, we can generate an
indication of how the GSMF behaves to masses lower than $10^{7.5} M_\odot$. For
the G10-COSMOS dataset, we use {\sc lambdar} photometric measurements of
approximately 170,000 galaxies compiled by \cite{Andrews2016}, along with a combination of
spectroscopic and photometric redshifts from \cite{Davies2015a}, taken
predominantly from \cite{Laigle2016}, and fit these galaxies with \magphys\ (as
we did with the GAMA equatorial sample, see \citealt{Driver2016b}). The
coverage in the G10 field of 1sqdeg is not nearly as high as in the equatorial
GAMA fields, but the sample extends $\sim 5$ mag fainter in the r-band.

\begin{figure*} \centering
%\vspace{3in}
%\includegraphics[scale=0.9]{FinalMassFunctions.pdf}
\includegraphics[scale=0.9]{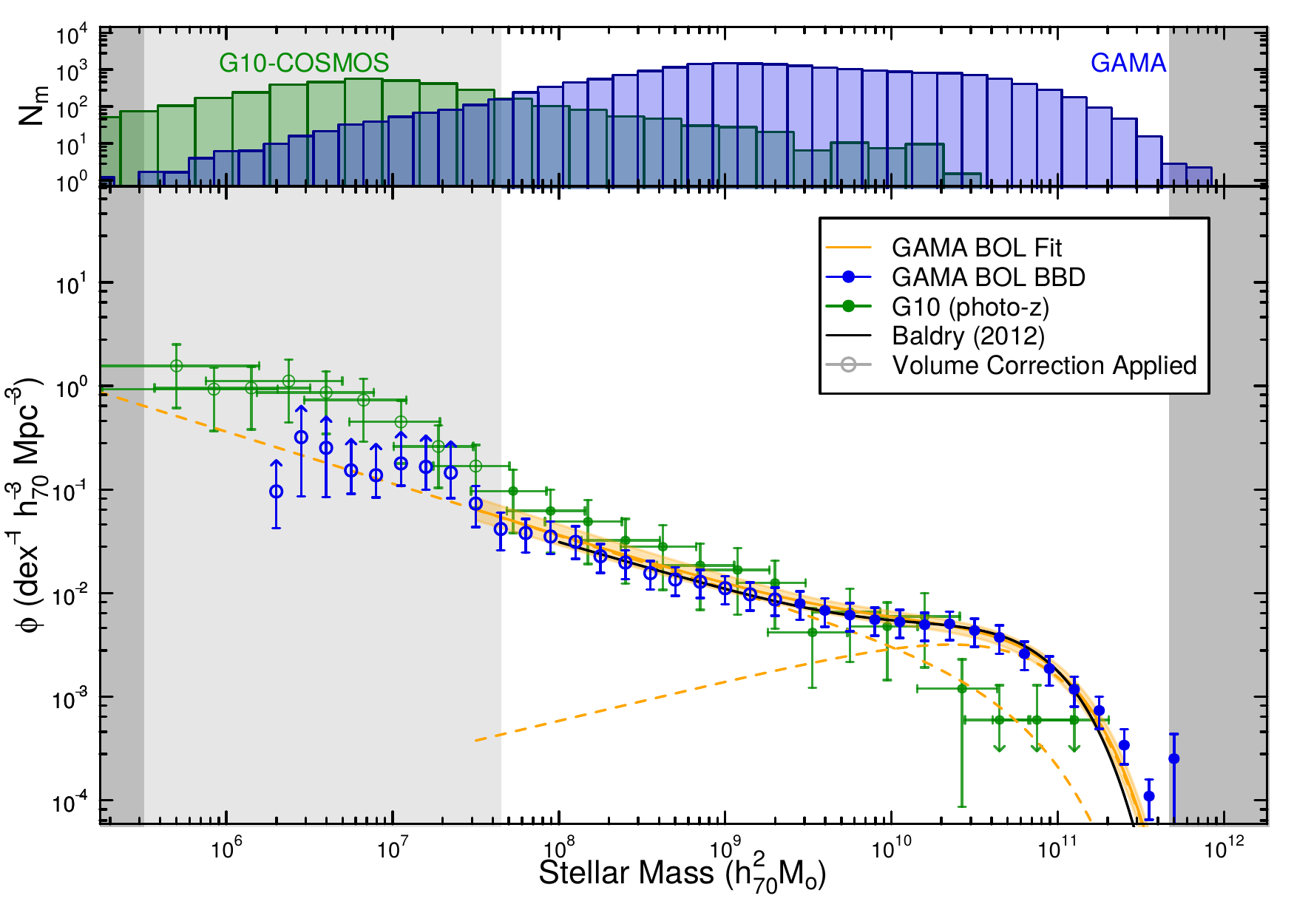}
\caption[The expanded GAMA +
G10-COSMOS GSMF]{The GSMF measured when combining data from GAMA with
that from G10-COSMOS. Uncertainty in the Mass direction for the G10-COSMOS binned data
indicate the median uncertainty in that bin caused {\em only} by photometric redshift uncertainties.
For this, we assume a standard $\Delta z_p = 0.02$.
From these uncertainties we can see that `bump' in number density coincides with where
Eddington bias becomes considerable, given the photometric redshift uncertainties.
The inclusion of the G10-COSMOS
allows us to probe to $10^{7.7}M_\odot$ without requiring any $V_{\rm max}$
corrections, and when incorporating corrections we are able to extend to masses as low as
$10^6 M_\odot$. Note, however, that we show bins requiring volume corrections with open circles, to
indicate that these bins may be biased by poorly determined redshift estimates, particularly in the G10-COSMOS.
Note also that the G10-COSMOS GSMF includes uncertainty due to cosmic variance using the estimator
from \protect \cite{Driver2010}. The light and dark grey regions in the figure indicate the masses
where we believe GAMA is systematically incomplete, and where both samples have low number statistics, respectively.
}\label{fig: GSMFs2} \end{figure*}

Using this combined sample, we are able to construct an indicative
GSMF to much lower masses than can be probed by GAMA alone. We construct a
simple binned GSMF for the G10 sample, without any attempt to match
samples or normalisation to those in GAMA. This combined dataset is
shown in Figure \ref{fig: GSMFs2}. Data for the binned GAMA GSMF shown in the figure
is {\bbf provided as a machine readable file alongside this work.}

We see general agreement between the extrapolated best-fit GAMA GSMF and the G10-COSMOS
sample down to masses as low as $10^6 M_\odot$, with the exception of a modest bump in
the faint end slope seen around $10^7 M_\odot$. This bump is likely to arise from Eddington
bias induced by the large stellar mass uncertainties, {\bbf however the similar rise in the
low mass tail of the GAMA BBD is a tantalising suggestion of, perhaps, a slight rise in the
faint end slope of the mass function. }
Nonetheless, this function rejoins
our extrapolation at $10^6 M_\odot$, {\bbf nominally below where we expect incompleteness to be problematic
in the COSMOS field,} and so we conclude that {\bbf (for now)} there appears to be no sign of any major up- or
down-turn to this limit.

Naturally this comparison is a qualitative rather than quantitative measure.
Nonetheless the agreement between the datasets in the range of overlap is good,
and overall provides a glimpse into the very low mass population and
that extrapolation to $10^6 M_\odot$ is not unreasonable.

%\begin{table}
%  \centering
%  \caption[Bin values of the GAMA BBD GSMF]{Values derived for the
%GAMA bolometric  GSMF from our BBD analysis. Here we show the binned number density and associated
%uncertainties after our Monte Carlo.}\label{tab: gamavalues} \begin{tabular}{c|cc}
%Survey      & bin centre & $\log_{10}(\phi)$  \\
%            & $\log_{10}(M_\star/M_\odot)$ & $\log_{10}({\rm dex}^{-1}{\rm Mpc}^{-3})$ \\
%\hline
%GAMA & 11.75 & $-4.004^{+0.048}_{-0.190}$ \\
%GAMA & 11.50 & $-3.648^{+0.052}_{-0.048}$ \\
%GAMA & 11.25 & $-3.157^{+0.049}_{-0.039}$ \\
%GAMA & 11.00 & $-2.805^{+0.027}_{-0.018}$ \\
%GAMA & 10.75 & $-2.563^{+0.015}_{-0.012}$ \\
%GAMA & 10.50 & $-2.395^{+0.011}_{-0.010}$ \\
%GAMA & 10.25 & $-2.343^{+0.019}_{-0.012}$ \\
%GAMA & 10.00 & $-2.312^{+0.009}_{-0.009}$ \\
%GAMA &  9.75 & $-2.247^{+0.009}_{-0.007}$ \\
%GAMA &  9.50 & $-2.173^{+0.008}_{-0.008}$ \\
%GAMA &  9.25 & $-2.095^{+0.008}_{-0.008}$ \\
%GAMA &  9.00 & $-2.011^{+0.012}_{-0.011}$ \\
%GAMA &  8.75 & $-1.851^{+0.015}_{-0.016}$ \\
%GAMA &  8.50 & $-1.809^{+0.018}_{-0.016}$ \\
%GAMA &  8.25 & $-1.599^{+0.033}_{-0.032}$ \\
%GAMA &  8.00 & $-1.402^{+0.050}_{-0.041}$ \\
%GAMA &  7.75 & $-1.327^{+0.070}_{-0.064}$ \\
%GAMA &  7.50 & $-1.168^{+0.137}_{-0.108}$ \\
%GAMA &  7.25 & $-0.946^{+0.179}_{-0.160}$ \\
%GAMA &  7.00 & $-0.962^{+\infty}_{-0.208}$ \\
%GAMA &  6.75 & $-0.924^{+\infty}_{-0.330}$ \\
%GAMA &  6.50 & $-1.106^{+\infty}_{-\infty}$ \\
%\hline \end{tabular} \end{table}

%}}}
\section{Contribution to $\Omega_\star$} %{{{
\begin{figure*} \centering
%\vspace{3in}
%\includegraphics[scale=0.9]{GSMF+Sims.pdf}
\includegraphics[scale=0.9]{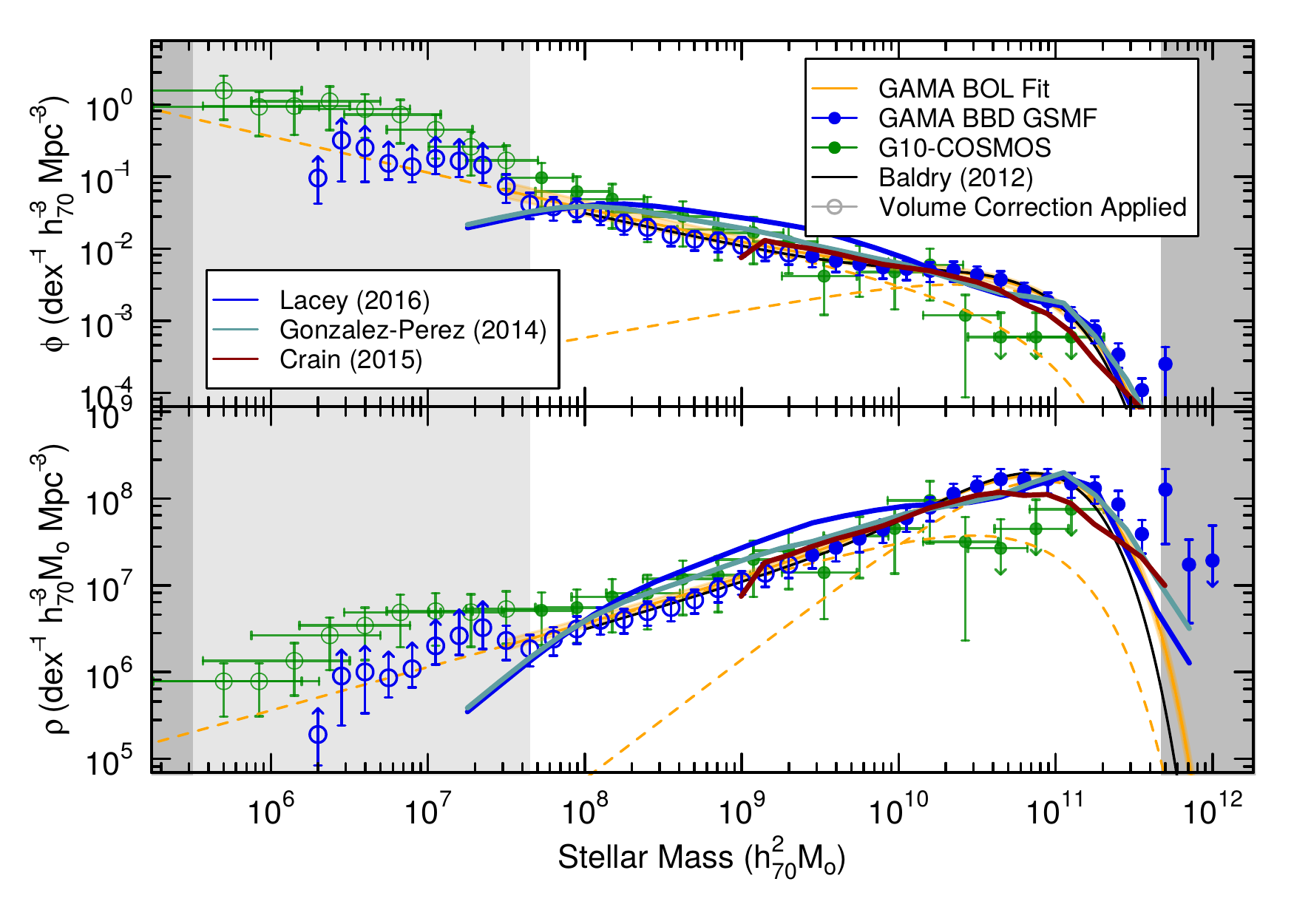}
\caption[Various stellar mass
function distributions compared to simulations]{Our measured stellar mass
function, shown as number density and mass density, compared to those measured
in GALFORM semi-analytic models and EAGLE hydrodynamic simulations. The bin uncertainties
in the mass-direction show the median mass uncertainty for sources in that bin which
are caused {\em only} from (an assumed typical) standard photometric redshift uncertainty $\Delta z_p = 0.02$.
Note that the G10-COSMOS binned GSMF, and the GAMA DCMV GSMF fit, both include their
respective uncertainties due to cosmic variance using the estimator from \protect \cite{Driver2010} in the
number density direction. The light and dark grey regions in the figure indicate the masses
where we believe GAMA is systematically incomplete, and where both samples have low number statistics, respectively.
}\label{fig: GSMF distr} \end{figure*}

To conclude, we can utilise
our fitted GSMF to derive the value of the stellar mass density parameter
$\Omega_\star$ and the fractional contribution of stars to the universal baryon
density $\Omega_b$. Furthermore, we can be somewhat confident in extrapolating our fit
down to much lower masses than GAMA alone would allow, given the consistency we
see in the GAMA+G10-COSMOS GSMF. Figure \ref{fig: GSMF distr} shows the
distributions of stellar mass number density $\phi$, and mass density $\rho$, for
our final GSMF. In the figure we also compare these distributions to those from
the GALFORM semi-analytic models of \cite{Lacey2016,Gonzalez-Perez2014}, and to
the hydrodynamic simulations from EAGLE \cite{Schaye2015,Crain2015}.

From the mass density distribution in Figure \ref{fig: GSMF distr}, we can see
that the stellar mass density is dominated by ${\mathcal M}^\star$ galaxies, as
has long been known. Our distributions match exceptionally well with the simulations, although this
is somewhat by design given that the GALFORM semi-analytic models are calibrated to the
B$_j$- and K-band luminosity functions at $z=0$ \citep{Lacey2016,Gonzalez-Perez2014}. We find a final
$\Omega_\star= 1.66^{+0.24}_{-0.23}\pm0.97  h^{-1}_{70} \times 10^{-3}$, corresponding to an overall percentage of
baryons stored in bound stellar material $f_b = 6.99^{+1.01}_{-0.97}\pm 4.09 $ (assuming the Planck
$\Omega_b=23.76\times10^{-3}h^{-2}_{70}$), inclusive of uncertainty due to cosmic variance and systematic
uncertainties in SPS modelling.

\begin{figure*} \centering
%\vspace{3in}
%\includegraphics[scale=0.8]{OmegaStar.pdf}
\includegraphics[scale=0.8]{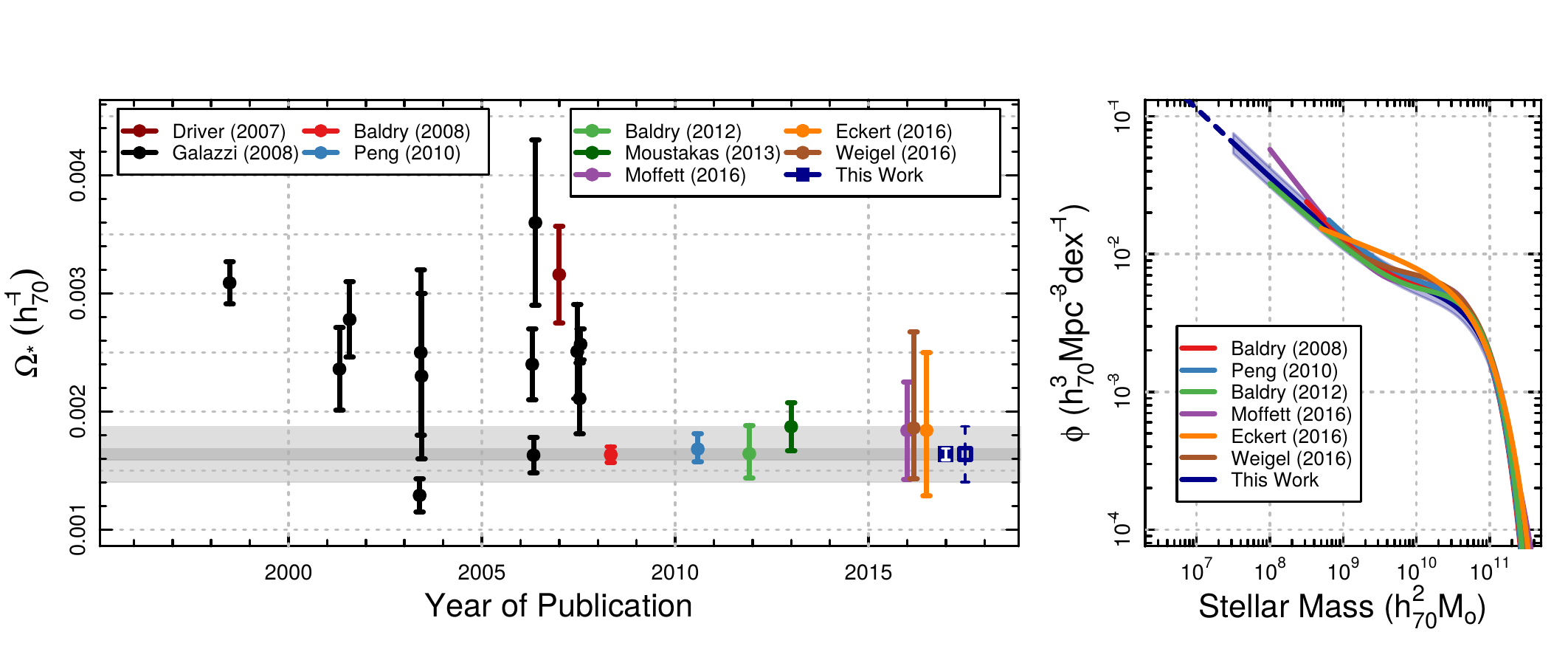}
\caption[Contribution to
$\Omega_\star$]{{\em Left:} Contribution of the full GAMA sample to
$\Omega_\star$, compared to previous estimates from the literature. The two blue data points
from this work correspond to estimates incorporating only the formal uncertainty on the fit
(solid symbol, dark grey shaded bar), and that including uncertainty due to cosmic variance (open
symbol, light grey shaded bar). Note the fit uncertainty is smaller than the data point in the former case, and therefore
has been shown as a white bar within the data point. {\bbf No data displayed here includes systematic uncertainty
due to stellar mass modelling and,} with the exception of our open symbol data point,
no measurement uncertainties displayed here include uncertainty due to cosmic variance. {\em Right:}
the GSMFs corresponding to the most recent estimates of $\Omega_\star$ in the left panel. This provides an indication
of the level of concordance in the literature with regards to the overall shape of the GSMF, and also of the
level of variation in the GSMF required to create a significant change in the value of $\Omega_\star$.
}\label{fig: omega star} \end{figure*}

{\bbf With respect to random uncertainties only, our estimate represents the
most stringent constraint on the bound component of both $\Omega_\star$ and $f_b$ to date.
As expected, however, our estimates
of both $\Omega_\star$ and $f_b$ are overwhelmingly dominated by the systematic
uncertainties in our mass estimation. Nonetheless, as these systematic uncertainties are inherently present in
all estimates of stellar masses, we can still perform an informative comparison between our value of
$\Omega_\star$ and a sample from the literature (seen in Figure \ref{fig: omega star}). This distribution
shows that, since 2008 there has been a reasonable consensus regarding the estimates of $\Omega_\star$.
This consensus is due, at least in part, by a consistent use of \cite{Bruzual2003} SPS models
in each of the post-2008 estimates (with the exception of that from \citealt{Moustakas2013}, who use a similar but
nonetheless different \citealt{Conroy2010} model), and to an increase
in sample sizes with the advent of big-data astronomy. In any case, the fact that all of
these estimates are subject to the same systematic uncertainties indicates
that, as a community, we are unlikely to gain further significant insight into the amount of mass stored in
bound stellar systems without: a significant reduction in the systematic uncertainties of stellar mass estimates,
a significant reduction in the masses of systems that we can analyse (see Section \ref{sec: discussion}), or both.

}

%}}}
\section{Conclusions}\label{sec: conclusions}%{{{
{\bbf We present the revised galaxy
stellar mass function for the GAMA $z\le 0.1$ sample, expanding on the GAMA-I analysis
presented in \cite{Baldry2012} to the full GAMA-II dataset in this volume. {\bbf We
utilise two stellar mass samples, calculated with and without consideration for the impact
of optically thick dust, finding no discernible difference between these samples.} As in
\cite{Baldry2012}, we calculate the GSMF using density-corrected maximum-volume (DCMV)
weights, defining our fiducial density using galaxies with $M_\star \ge 10^{10}
M_\odot$ in the redshift range $0.07 < z < 0.19$. Within these limits the cosmic
structure is fairly uniform, the sample is not yet affected by incompleteness,
and the volume is influenced by cosmic variance at the $<10\%$ level (using the
cosmic variance estimator from \citealt{Driver2010}), allowing for a stable
constraint on the fiducial average density. We fit the GSMF using a
Markov-Chain Monte-Carlo and mass limits defined
in a manner that is conservative with respect to incompleteness in both brightness and
colour. We choose to fit the GAMA low-z GSMF with a double
\cite{Schechter1976} function, finding best fit } parameters
$\mathcal M^\star=10^{10.78\pm0.01\pm0.20}M_\odot$,
$\phi^\star_1=(2.93\pm0.40)\times10^{-3}h_{70}^3$Mpc$^{-3}$,
$\alpha_1=-0.62\pm0.03\pm0.15$,
$\phi^\star_2=(0.63\pm0.10)\times10^{-3}h_{70}^3$Mpc$^{-3}$, and
$\alpha_2=-1.50\pm0.01\pm0.15$, {\bbf where the second uncertainty components on $\mathcal M^\star$ and each
$\alpha$ encode the systematic uncertainty on stellar mass estimation due to SPS modelling
uncertainties. The uncertainty due to cosmic variance is included in the stated uncertainties on
$\phi^\star_1$ and $\phi^\star_2$.} While the value
of ${\mathcal M}^\star$ here is higher than other works in the literature, we
argue that this is a result of the dedicated by-hand effort that was undertaken
to ensure photometry of the brightest systems in GAMA was accurately determined
\citep{Wright2016}.

{\bbf We explore the galaxy bivariate brightness distribution of stellar mass and
absolute surface brightness in order to explore the possible surface brightness incompleteness
of our dataset. Our BBD GSMFs both agree well with our nominal best-fit GSMFs from above,
however there is a slight excess in both mass samples at the lowest stellar masses. Furthermore,
the location of the known GAMA selection boundaries, and the distributions of known local sphere
galaxies from \cite{Karachentsev2004} and \cite{McConnachie2012}, both suggest that our sample may be
incomplete below $10^8$ in stellar mass.

To further explore the low-mass end of the GSMF,} we compare our estimated stellar mass function to the GSMF measured using
the same analysis applied to the G10-COSMOS dataset
\cite{Davies2015,Driver2016b,Andrews2016}. We find good agreement between the stellar mass
functions, and an indication that the faint end slope of the GSMF is relatively
well behaved down to masses as low as $M > 10^6 M_\odot$, showing an only
marginal feature at $\sim 10^7 M_\odot$.

We compare our measured mass function to those from the GALFORM semi-analytic
models \citep{Lacey2016,Gonzalez-Perez2014}, and to the GSMF from the EAGLE
hydrodynamic simulation \citep{Schaye2015,Crain2015}.
We find an exceptional agreement between the GALFORM semi-analytic models and our GSMF,
however this is arguably somewhat by design as the semi-analytic models are calibrated to the
B$_j$- and K-band luminosity functions at $z=0$ \citep{Lacey2016,Gonzalez-Perez2014}.

We compute the value of the stellar mass density parameter $\Omega_\star$ for
our mass function fit, finding $\Omega_\star= 1.66^{+0.24}_{-0.23}\pm0.97  h^{-1}_{70} \times 10^{-3}$,
corresponding to an overall
percentage of baryons stored in bound stellar material $f_b = 6.99^{+1.01}_{-0.97}\pm4.07$ (assuming the Planck
$\Omega_b=23.76\times10^{-3}h^{-2}_{70}$), inclusive of uncertainty due to cosmic variance and systematic uncertainties from
SPS modelling. Finally, using the joint
dataset from GAMA and G10-COSMOS, we conclude that there is no strong indication of a
significant up- or down-turn in the GSMF to stellar masses greater than $10^6 M_\odot$.
We conclude that the integrated stellar mass
density of bound material down to $M > 10^6 M_\odot$ is
well constrained, and that the fraction of universal baryonic matter stored in bound stellar material
within galaxies {\bbf (assuming our various SPS model parameters) } is unlikely to exceed $\sim8\%$.
{\bbf However, systematic uncertainties from the SPS models dominate our error-budget, and could possibly
drive this value as high as $\sim20\%$, assuming the most extreme SPS and IMF models. Additionally, }
the question of the amount of unbound stellar mass in halos remains open.
%}}}
\section*{Acknowledgements} %{{{
{\bbf We thank the referee, Eric Bell, for his thorough reading of our work and for his many constructive
suggestions. } GAMA is a joint European-Australasian project based around a spectroscopic campaign using
the Anglo-Australian Telescope. The GAMA input catalogue is based on data taken from the
Sloan Digital Sky Survey and the UKIRT Infrared Deep Sky Survey. Complementary imaging of the GAMA
regions is being obtained by a number of independent survey programmes including GALEX MIS, VST KiDS,
VISTA VIKING, WISE, Herschel-ATLAS, GMRT and ASKAP providing UV to radio coverage. GAMA is funded by
the STFC (UK), the ARC (Australia), the AAO, and the participating institutions.
The GAMA website is \url{http://www.gama-survey.org/}.
Based on observations made with ESO Telescopes at the La Silla Paranal Observatory
under programme ID 179.A-2004.
SJM acknowledges support from European Research Council Advanced Investigator Grant COSMICISM, 321302;
and the European Research Council Consolidator Grant {\sc CosmicDust} (ERC-2014-CoG-647939, PI H\,L\,Gomez).

%}}}
\bibliographystyle{mnras} \bibliography{GSMF2}
\appendix %{{{
\section{Fits without Fluxscale Correction}\label{appendix: nofluxcorr}
As discussed in Section \ref{sec: Two} the fluxscale parameter, while necessary,
may be affected by unrecognised systematic biases. Here we present the
GSMF and $\Omega_\star$ estimates determined when not incorporating the fluxscale
parameter. This provides a quasi-lower limit on our fits and parameter estimations,
and demonstrates the impact of this corrective factor.

\begin{figure*} \centering
\includegraphics[scale=0.9]{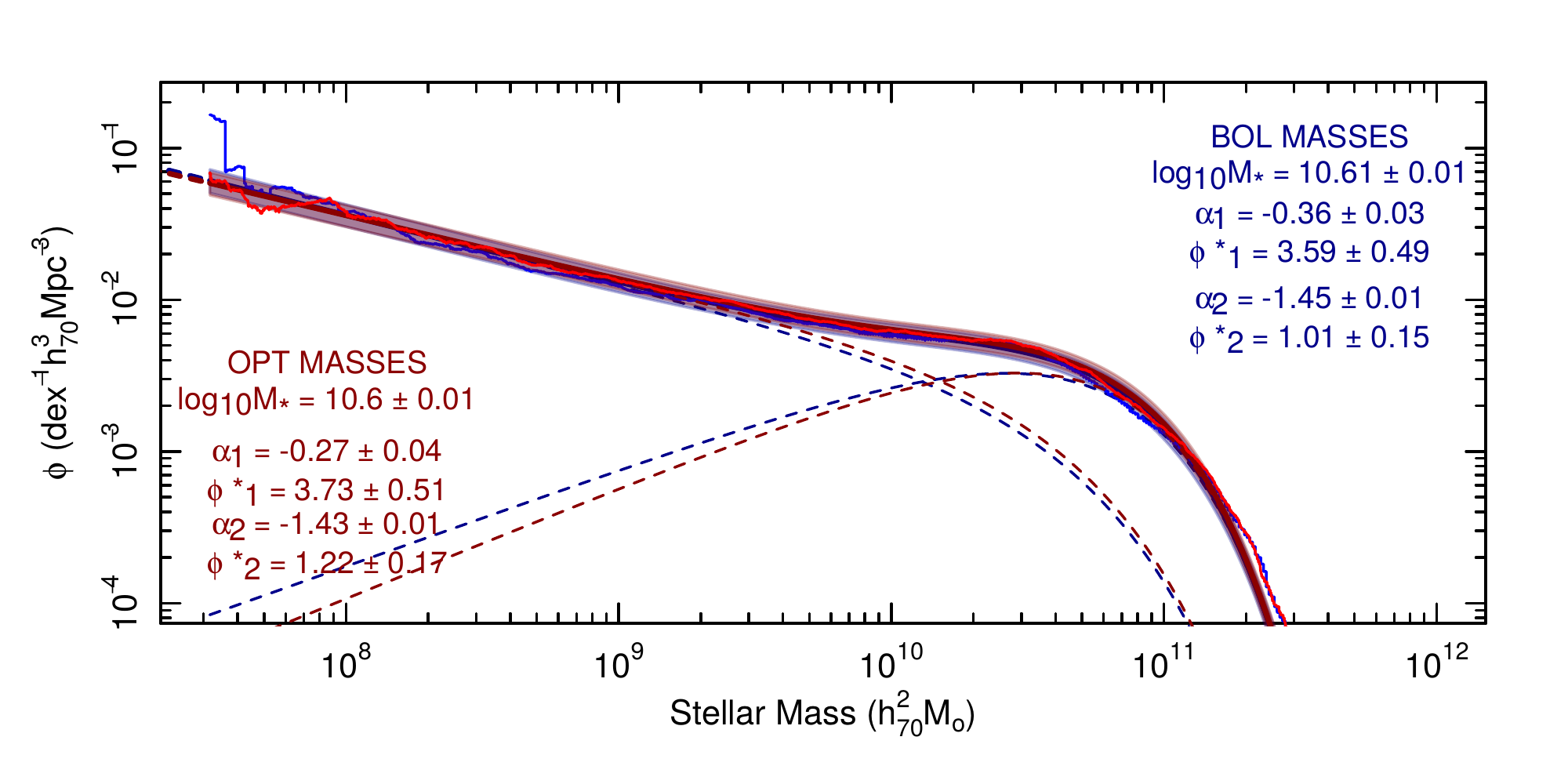}
\caption[The Best-Fit
GSMF for the GAMA low-z sample when not performing the fluxscale correction]{
The best-estimate GSMF for the GAMA low-z sample when not performing the fluxscale correction.
%The fit is estimated by applying a informative prior on the $\alpha_1$ parameter (a
%Gaussian with $\mu = -0.55$ and $\sigma=0.04$).
The figure annotations are the same as in Figure \ref{fig: gsmf}.
Note that these fits include uncertainty due to cosmic variance
using the estimator from \protect \cite{Driver2010}. Our standard
systematic uncertainties on ${\mathcal M}^\star$ and $\alpha$ are not shown.}\label{fig: GSMF nofluxscale}
\end{figure*}

\begin{figure*} \centering
%\vspace{3in}
%\includegraphics[scale=0.8]{OmegaStar_NoFluxScale.pdf}
\includegraphics[scale=0.8]{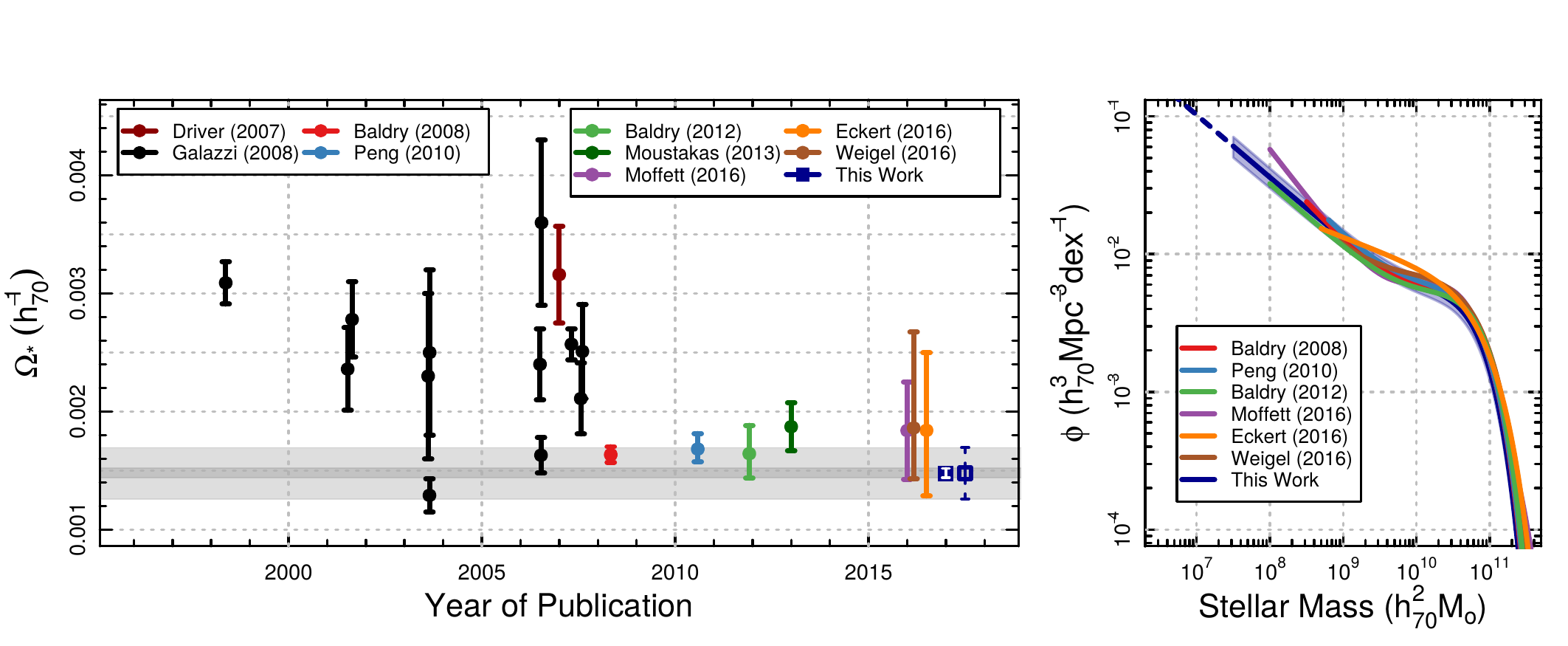}
\caption[Contribution to
$\Omega_\star$ without fluxscale]{{\em Left:} Contribution of the full GAMA sample to
$\Omega_\star$, compared to previous estimates from the literature, when not incorporating the fluxscale
correction. The figure is annotated as in Figure \ref{fig: omega star}. {\em Right:}
the GSMFs corresponding to the most recent estimates of $\Omega_\star$ in the left panel. This provides an indication
of the level of concordance in the literature with regards to the overall shape of the GSMF, and also of the
level of variation in the GSMF required to create a significant change in the value of $\Omega_\star$.
}\label{fig: OmegaStar nofluxscale} \end{figure*}

Figure \ref{fig: GSMF nofluxscale} shows the final GSMF estimated when not incorporating
the fluxscale parameter. It is the no-fluxscale equivalent of Figure \ref{fig: gsmf}.
Similarly, Figure \ref{fig: OmegaStar nofluxscale} shows the final estimate of $\Omega_\star$ when
not incorporating the fluxscale parameter.

In these figures, we can see that the most substantial change is in the extent of the GSMF
to the highest masses. This is not surprising as the fluxscale factor is expected to influence
high Sersic index sources the most, and these are overly contained at the highest mass end of the
sample (\eg elliptical sources and bulge-dominated disks). The result is that our sample loses
a substantial amount of mass in the same region where the mass-density function peaks. This drives the
significant loss in the stellar mass density parameter.


\section{Fits with decoupled ${\mathcal M}^\star$}\label{appendix: decoupled}
\begin{figure*} \centering
\includegraphics[scale=0.9]{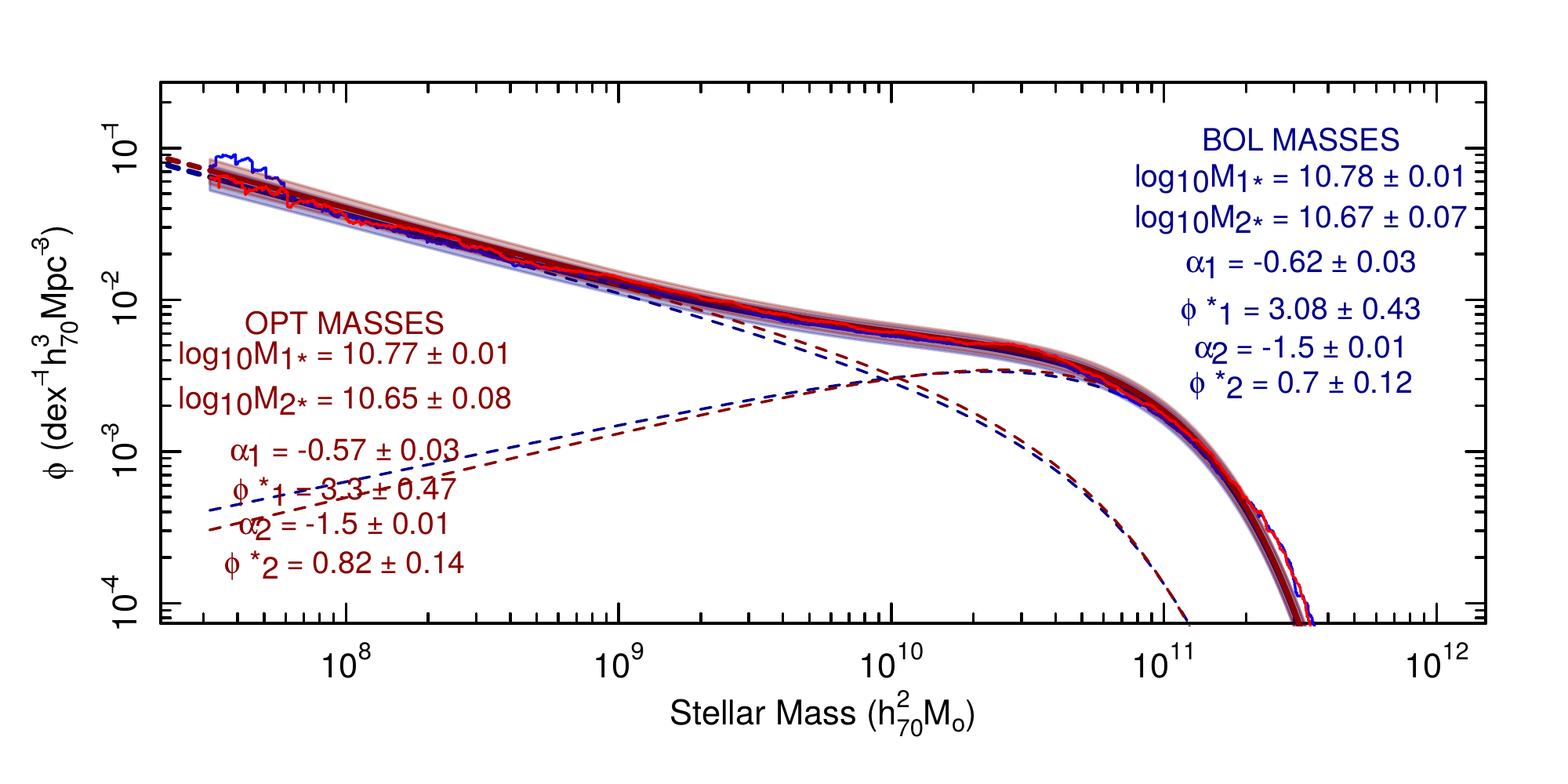}
\caption[The Best-Fit
GSMF for the GAMA low-z sample when fitting with decoupled ${\mathcal M}^\star$ parameters]{
The best-estimate GSMF for the GAMA low-z sample when fitting with decoupled ${\mathcal M}^\star$ parameters.
%The fit is estimated by applying a informative prior on the $\alpha_1$ parameter (a
%Gaussian with $\mu = -0.55$ and $\sigma=0.04$).
The figure annotations are the same as in Figure \ref{fig: gsmf}.
Note that these fits include uncertainty due to cosmic variance
using the estimator from \protect \cite{Driver2010}.  Our standard
systematic uncertainties on ${\mathcal M}^\star$ and $\alpha$ are not shown.}\label{fig: decoupled}
\end{figure*}

{\bbf As discussed in Section \ref{sec: Three}, we opt to fit our main GSMFs with a coupled
${\mathcal M}^\star$ 2-component Schecter function. This choice is motivated mostly to enable
simple comparison with previous GSMF fits. However extensive work in exploring individual populations
of galaxies separated by morphology and dynamical properties \citep{Moffett2016, Kelvin2014}, demonstrate that
many decoupled Schechter functions are required to capture the true diversity of galaxy mass functions. With this in mind,
we briefly explore the fits obtained when using a decoupled 2-component Schechter function, in Figure
\ref{fig: decoupled}.

The fit parameters from our decoupled fit indicate that data prefers a decoupled ${\mathcal M}^\star$ only
slightly. The two free ${\mathcal M}^\star$ parameters end up with values that are only slightly
inconsistent with each other, and otherwise the fit parameters are largely unchanged from our original
coupled fits. Nonetheless, the fact that the lower-mass component favours a slightly lower ${\mathcal M}^\star$
than the higher mass component is consistent with the results found previously in the
literature, such as previously in the GAMA low-z sample by \cite{Moffett2016}.}

\section{Deriving Mass Limits}\label{appendix: mass limits}
\begin{figure*} \centering
%\vspace{3in}
%\includegraphics[scale=0.3]{MassLimits1.png}
\includegraphics[scale=0.5]{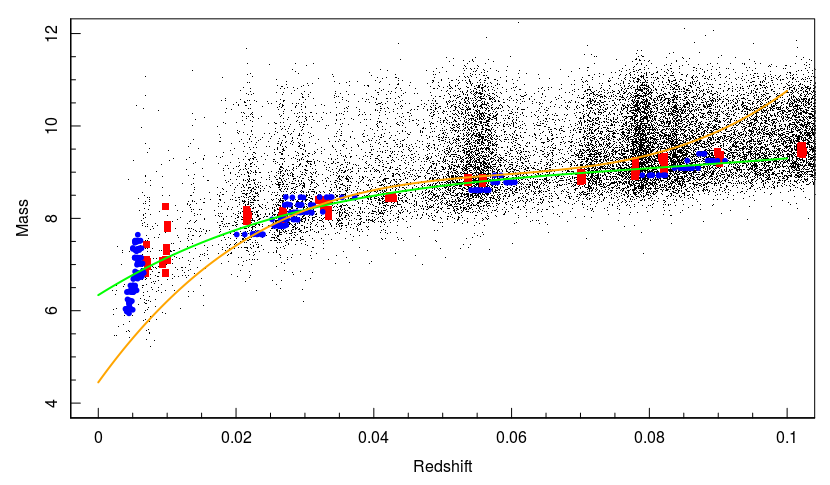}
\caption[Mass Limits returned by our automated procedure]{
Demonstration of the stellar mass limits returned by our automated mass limit estimation procedure. Here we can see the
distribution of bootstrapped turn-over estimates, derived in comoving distance bins (red) and mass bins (blue).
These turn over estimates are then fit with a polynomial (green). For comparison, the mass limit function of
\protect \cite{Moffett2016} is shown in orange. Note that this is a generic diagnostic figure output by the function,
and therefore is intentionally not created with meaningful axis labels. }\label{fig: booted masslim}
\end{figure*}

\begin{figure*} \centering
%\vspace{3in}
%\includegraphics[scale=0.3]{MassLimits1.png}
\includegraphics[scale=0.2]{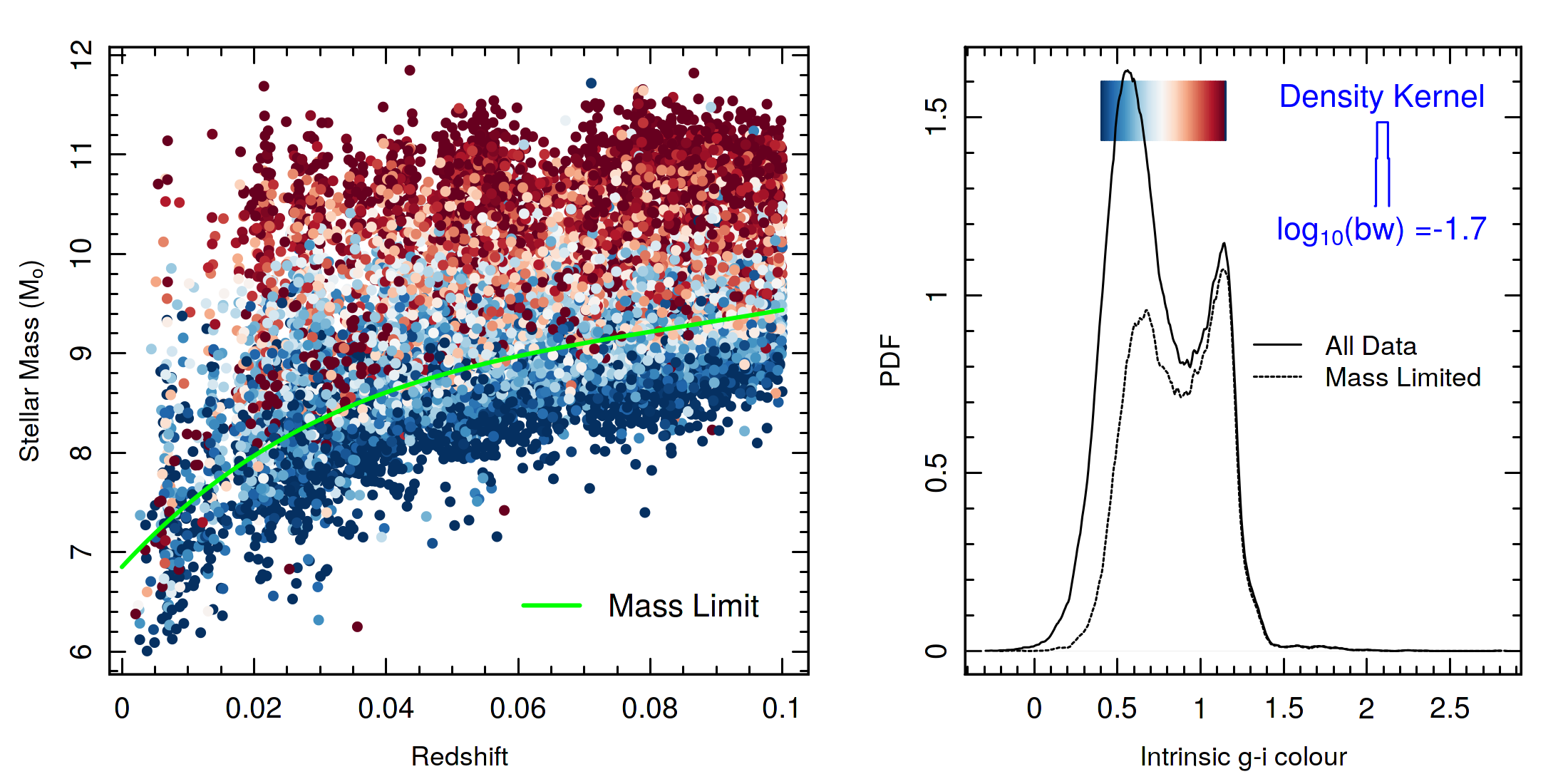}
\caption[Mass Limits and colour, showing colour debiasing]{
{\bbf Demonstration that the stellar mass limits returned by our automated mass limit estimation procedure adequately debias the colour
distribution of galaxies within our sample. The left-hand figure shows the mass-redshift space of all galaxies within our sample,
coloured by g-i colour, along with the mass limit function. The distribution of colours (and the colour-bar) is shown in the right
hand panel. The solid line in the right hand figure is the colour distribution of the full sample, and the dashed line shows the
distribution after applying the mass-limit cut. Note in-particular that the mass-limit cut preferentially removes blue galaxies, which
are visible to higher redshifts (at a given stellar mass) than their red counterparts.}}\label{fig: masslim colour}
\end{figure*}

For the automated derivation of mass limits, we fit a polynomial to bootstrapped
estimates of the turn-over point of the comoving galaxy number density as a function
of stellar mass, and of the turn over of the stellar mass density as a function of
comoving distance.
The result of this procedure is shown in Figure \ref{fig: booted masslim}, where
we show the individually estimated turn-over points in each dimension. These points
have then been fit by a polynomial, yielding the mass limit function shown. {\bbf Importantly,
Figure \ref{fig: masslim colour} demonstrates that the mass limits successfully debias
the sample with respect to colour, as seen by the mass limit preferentially removing blue galaxies
(which are visible to higher redshifts than their red counterparts).}

\begin{figure*} \centering
%\vspace{3in}
%\includegraphics[scale=0.4]{MassLimitsComparison2.png}
%\includegraphics[scale=0.4]{MassLimitsComparison1.png}
\includegraphics[scale=0.2]{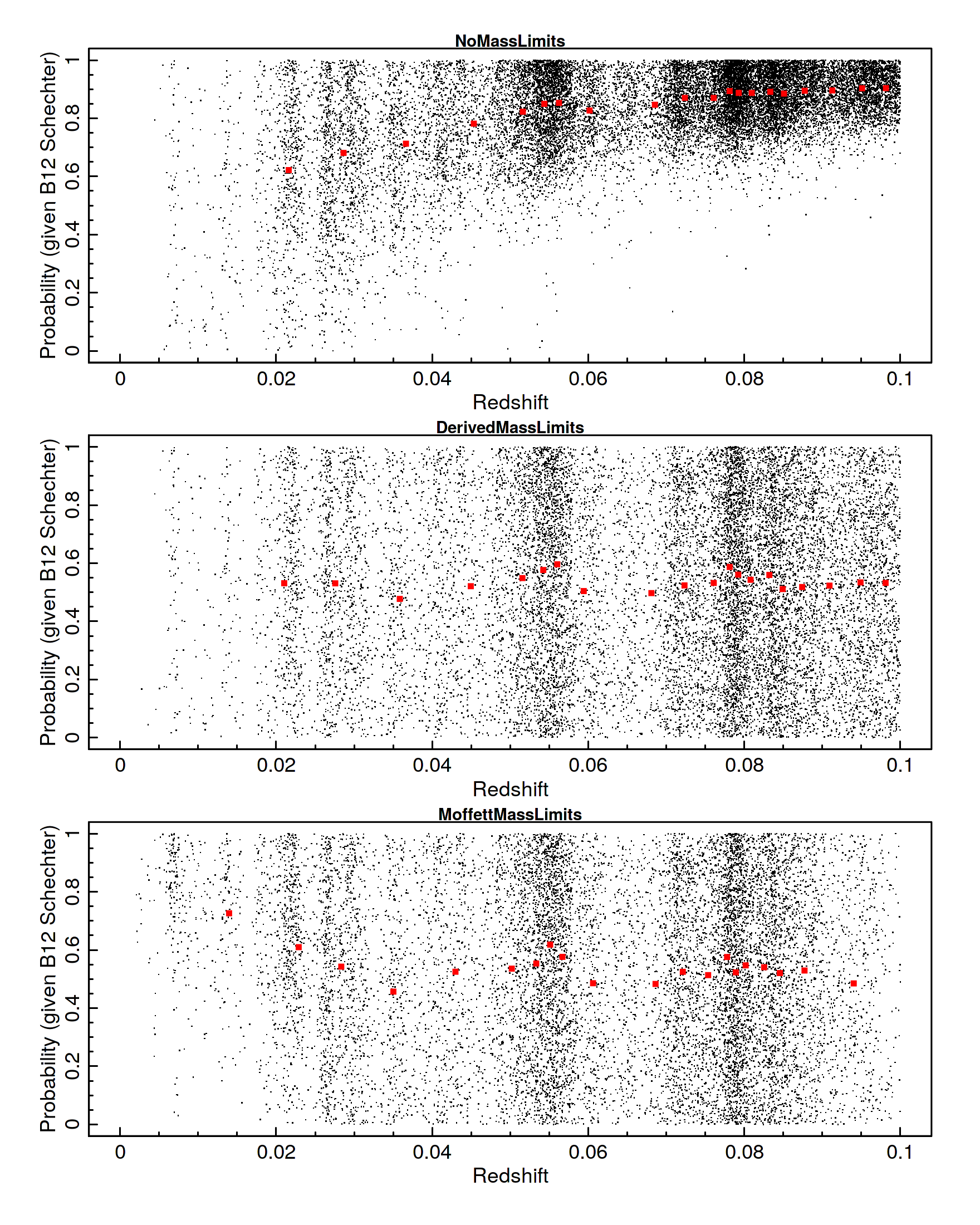}%\includegraphics[scale=0.4]{FigureC3b.png}
\caption[Comparison between derived mass limits]{
  Comparison between the galaxy probability distribution assuming: ({\em top}) constant mass limits, ({\em middle}) those derived
  using our automated procedure, and ({\em bottom}) mass limits presented in \protect \cite{Moffett2016}.
  We see that the \protect \cite{Moffett2016} mass limits show a deviation away from
  the expectation probability (red points) of 0.5 at low masses, indicating that the mass limit there is not accurate (assuming,
of course, that the GSMF is reasonably represented by the \citealt{Baldry2012} GSMF). Conversely, the automatically defined
mass limits returned from our procedure show no such bias.}\label{fig: masslim compar}
\end{figure*}

We then
determine the fidelity of these mass limits by comparing the distribution of the mass-limited
galaxy probability function (with redshift) when using these mass limits and the mass limits
implemented in \cite{Moffett2016}. To do this, we assume a \cite{Baldry2012} double Schechter
function and compute the probability of observing each galaxy given this GSMF and the assigned
mass limit. The distribution of probabilities using these two mass limit functions are given in
Figure  \ref{fig: masslim compar}. {\bbf These figures show the distribution of galaxy probabilities
assuming a \cite{Baldry2012} generative distribution and the relevant mass-limit function. If both of
these are a good reflection of the data, the distribution of probabilities should have an expectation of
$0.5$. In these figures we can see that the \cite{Moffett2016}
mass limits are systematically biased at low redshift, indicated by an expectation systematically different
from $0.5$}. Conversely, we see that the automatically defined mass limits show no such systematic bias.

%}}}

\end{document}